\documentclass[sigconf]{acmart}

\renewcommand\footnotetextcopyrightpermission[1]{} 
\usepackage{xcolor, tcolorbox, multirow, multicol, dcolumn, threeparttable, color, subfig, graphicx, mathrsfs, amsmath}

\AtBeginDocument{%
  \providecommand\BibTeX{{%
    \normalfont B\kern-0.5em{\scshape i\kern-0.25em b}\kern-0.8em\TeX}}}

\begin{document}
\title[Who, What, Why and How? Towards the Monetary Incentive\dots]{Who, What, Why and How? Towards the Monetary Incentive in Crowd Collaboration: A Case Study of Github's Sponsor Mechanism}

\author{Xunhui Zhang, Tao Wang*, Yue Yu*, Qiubing Zeng, Zhixing Li, Huaimin Wang}
\email{{zhangxunhui, taowang2005, yuyue, lizhixing15}@nudt.edu.cn, qiubingzeng@gmail.com, whm\_w@163.com}
\affiliation{%
  \institution{National University of Defense Technology}
  \streetaddress{No. 109, Deya Road}
  \city{Changsha}
  \state{Hunan}
  \country{China}
  \postcode{410073}
}

\thanks{* indicates the corresponding authors}

\renewcommand{\shortauthors}{Zhang et al.}

\begin{abstract}
  While many forms of financial support are currently available, there are still many complaints about inadequate financing from software maintainers. In May 2019, GitHub, the world's most active social coding platform, launched the \textsc{Sponsor} mechanism as a step toward more deeply integrating open source development and financial support.
  This paper collects data on 8,028 \emph{maintainers}, 13,555 \emph{sponsors}, and 22,515 \emph{sponsorships} and conducts a comprehensive analysis. We explore the relationship between the \textsc{Sponsor} mechanism and developers along four dimensions using a combination of qualitative and quantitative analysis, examining \textbf{why} developers participate, \textbf{how} the mechanism affects developer activity, \textbf{who} obtains more sponsorships, and \textbf{what} mechanism flaws developers have encountered in the process of using it.
  We find a long-tail effect in the act of sponsorship, with most maintainers' expectations remaining unmet, and sponsorship has only a short-term, slightly positive impact on development activity but is not sustainable. While sponsors participate in this mechanism mainly as a means of thanking the developers of OSS that they use, in practice, the social status of developers is the primary influence on the number of sponsorships.
  We find that both the \textsc{Sponsor} mechanism and open source donations have certain shortcomings and need further improvements to attract more participants.
\end{abstract}


\keywords{sponsor, donation, GitHub, open source, financial support}

\maketitle

\section{Introduction}









Open source development has brought prosperity to software ecosystems. Its characteristics of distributed coordination, free participation, and convenient sharing have led to the emergence of myriad open source projects, 
large-scale participation of developers, and continuous development of high-quality projects.
However, the expansion of project scales has also brought challenges for software maintenance, such as continuously and rapidly increasing feature requests and bug fix reports \cite{Jaweria-bug-prior} and an increasing pull request review workload \cite{Jing-challenges}.
Although there are many continuous integration (CI) tools and continuous deployment (CD) tools to help reduce the workload of project managers, the complicated and high-pressure maintenance work still subjects them to stress \cite{Tung-Sanfilippo}.

Past studies have shown that most current open source work is still spontaneously performed by volunteers \cite{open-source-survey}. They engage in open source work as a hobby, to improve their personal reputations or to learn new technologies. These intrinsic benefits motivate volunteers to make open source contributions \cite{open-source-get-paid}.
However, many core managers and software maintainers would like to secure funding from others for their open source work because of the aforementioned challenges, thereby alleviating the related mental pressure and financial burdens \cite{Steven-hard-work, Schlueter-money, The-ethics}.

At present, there are many ways in which the open source sphere obtains financial support, such as crowdfunding on Kickstarter, project donations on OpenCollective, and issue rewards on BountySource and IssueHunt \cite{nayafia-financial}.
However, these are mainly web portals serving open source contributors active in other social coding communities.
The separation of development activities and financial support brings problems. First, it is difficult for sponsors to find active developers and open source projects in the open source community. Second, open source contributors need to spend considerable effort on maintaining the financial support platform.
In May 2019, GitHub, the world's most popular software hosting platform, launched the \textsc{Sponsor} mechanism, characterized by deep integration of financial support and the social coding platform.
While the \textsc{Sponsor} mechanism supports sponsorship of organizations and projects, it targets mainly individual contributors in the GitHub community. Therefore, unlike past related studies \cite{Overney-hanging, Overney-rich}, we can explore donation mechanism in the open source sphere from the perspective of individual developers.
In this context, this paper aims to \emph{explore donation in the open source sphere using the \textsc{Sponsor} mechanism as an example}.
We conducted an empirical study based on mixed methods and answered the following research questions.
\begin{enumerate}
  \renewcommand{\labelenumi}{\textbf{RQ\theenumi}}
\item \textbf{\textit{Why} do individuals participate or not in the \textsc{Sponsor} mechanism?}

From the feedback of GitHub developers, we summarized eight reasons for participation among sponsored developers, six reasons for participation among sponsors, and six reasons for not participating in the mechanism among other individuals. 
The main reason that participants used the \textsc{Sponsor} mechanism was its relationship with open source software (OSS) usage. The main reason for not participating was that developers did not need sponsorship or that they were driven to participate in open source development because of its nonmonetary character. 
Our findings can help optimize the \textsc{Sponsor} mechanism and attract more participants by satisfying the different motivations of contributors.

\item \textbf{\textit{How} effective is sponsorship in motivating developer OSS activity?}

We find through quantitative analysis that the sponsor mechanism has provided only a short-term, subtle boost to contributors' activities. According to the results of the qualitative analysis, most developers agree that sponsorship can provide them with motivation but are not satisfied with the available amounts. In contrast, most sponsors are satisfied with the current mechanism.
Our findings shed light on the application of the \textsc{Sponsor} mechanism in the open source sphere and the problems surrounding it. This work 
helps to rationalize the mechanism to promote greater participation in open source contributions among developers.

\item \textbf{\textit{Who} is likely to receive more sponsorship?}

The questionnaire results show that making useful OSS contributions and being active are the most critical factors for obtaining more sponsorship. However, according to the quantitative data analysis results, the factor that most affects sponsorship is the developer's social status in the community.
Our findings can provide actionable suggestions for developers seeking more sponsorships, while the conflicting results also illuminate the problems with OSS donations.

\item \textbf{\textit{What} are the shortcomings of the \textsc{Sponsor} mechanism?}

The research reveals that problems with the mechanism include usage deficiencies, object orientation with supported functions, and personalization. Many developers complain that the donations do not apply to open source ecosystems. A more relevant mechanism is needed to promote the healthy and sustainable development of the ecosystem.
\end{enumerate}

The contributions of this paper are as follows:
\begin{itemize}
\item To the best of our knowledge, this is the first in-depth study that comprehensively analyzes the GitHub \textsc{Sponsor} mechanism.
\item We quantitatively and qualitatively analyze the \textsc{Sponsor} mechanism along four dimensions, including developers' motivation to participate (why), the mechanism's effectiveness (how), the characteristics of developers who obtain more sponsorships (who), and the mechanism's shortcomings (what).
\item We provide actionable suggestions to help developers participating in the \textsc{Sponsor} mechanism obtain more sponsorship and feasible advice for improving the mechanism's effectiveness.
\end{itemize}

The remainder of this paper is organized as follows. 
Section \ref{related-work} presents the related work, and
Section \ref{background} describes the background of the GitHub \textsc{Sponsor} mechanism. Section \ref{method} presents the study design of this paper. In Section \ref{results}, we describe the results for each research question. Then, we discuss the findings in Section \ref{discussion}, and describe the threats in Section \ref{threats}. Finally, in Section \ref{conclusion}, we conclude the paper and describe future work.

\section{Related Work}
\label{related-work}

Open Innovation in Science (OIS) is a concept, which unifies the two domains of open and collaborative practices in science, \emph{i.e.,} open science (OS) and open innovation (OI)~\cite{beck2020open}.
For OS, the three pillars are accessibility, transparency, and inclusivity, among which the inclusivity (\emph{e.g.,} citizen science) is directly related to the knowledge production process.
For OI, various forms of collaborative practice exist, including crowdsourcing, OSS development, etc.
Regarding these open initiatives, the motivation and incentives of participation has always been the focus of continuous research~\cite{willinsky_unacknowledged_2005,antikainen2010rewarding}.
Although there are different views on the relationship between citizen science, crowdsourcing, and OSS development, we follow the relationships described above and present the related work on participation motivation and monetary incentives of the three parts separately.

\subsection{Citizen science}
For traditional citizen science, the motivation of participants varies greatly depending on the age~\cite{alender2016understanding}, gender~\cite{mesch2006effects}, educational background~\cite{macphail2020power}, and level of involvement~\cite{tiago2017influence}.
In many cases, both monetary and non-monetary incentives have a positive effect on participation~\cite{cappa2018bring}.
However, Wiseman et al. found that non-monetary incentives alone were better for online HCI projects to promote high-quality data from participants~\cite{wiseman2017exploring}.
Knowles~\cite{knowles2013cyber} also confirmed that although monetary incentives enhanced participation, they undermined sustained participation in volunteering initiatives. While for some specific projects (\emph{e.g.,} the conservation of species), monetary incentives even have the opposite effect~\cite{RICHTER2021109325}.

Because participants act as sensors to collect data or volunteer their idling computer or brainpower to classify large data sets in the citizen science projects~\cite{wiseman2017exploring}, their motivation to participate is primarily intrinsic~\cite{larson2020diverse,domroese2017watch}. However, as motivation to participate varies for different projects, the imposition of monetary incentives can have different effects. Unlike traditional citizen science, OSS development is an open innovation activity requiring deep involvement and a great deal of experience, so the motivation and incentives for participation may vary considerably.

\subsection{Crowdsourcing}
Acting as a type of online activity, participants will receive the satisfaction of a given kind of need, be it economic, social recognition, self-esteem, or the development of individual skills~\cite{estelles2012towards}.
Hossain~\cite{hossain2012users} classified the motivators into extrinsic and intrinsic motivators, where extrinsic motivators include financial motivators (\emph{e.g., cash}), social motivators (\emph{e.g., peer recognition}), and organizational motivators (\emph{e.g., career development}). Intrinsic motivators are directly related to participants' satisfaction with the task (\emph{e.g., enjoyment, fun}).
Considering the related incentives,
Liang et al.~\cite{liang2018intrinsic} highlighted that both intrinsic and extrinsic incentives could increase the effort of participation; however, extrinsic incentives weaken the impact of intrinsic motivation.
By comparing paid and unpaid tasks, Mao et al.~\cite{mao2013volunteering} concluded that monetary incentives make the task processing speed faster, but the quality is reduced.
Based on this, Feyisetan et al.~\cite{feyisetan2015improving} improved the paid microtasks more engaging by including sociality features or other game elements.
MTurk is a typical and popular crowdsourcing platform based on financial incentives and gamification, where participants are recruited, paid, and rated for their participation in microtasks, which ensure speed and quality at the same time~\cite{casler2013separate}.
Unlike MTurk, the contribution to Wikipedia is not incentivized by monetary rewards. Content contribution is more driven by reciprocity, self-development, while community participation relies on altruism, self-belonging, etc~\cite{xu2015empirical}.

As can be seen from the related works above, there are many situations of crowdsourcing and different forms of motivation and incentive. However, unlike OSS development, traditional crowdsourcing tasks are mostly micro-tasks, which are relatively simple and require less time. Moreover, there is a clear distinction between the roles, \emph{i.e.,} core developers and external contributors for OSS contributors. Contribution types include code contribution, code review, repository maintenance, management, etc.

\subsection{Open source software development}

Successful OSS initiatives can effectively change the method of software development~\cite{Gutwin-group, Kogut-distributed}, improve software development efficiency~\cite{Haefliger-reuse, Sojer-reuse}, and ensure software quality through effective management~\cite{Aberdour-quality, Schmidt-quality}.
Many projects have emerged along with the increasing number of users participating in the development of the OSS community~\cite{Github-report}.
In this context, many companies are involved in contributing to open source projects~\cite{Harvey-company}. However, they have limited control and influence in day-to-day OSS work and decision processes~\cite{Izquierdo-company}, and OSS still relies on the voluntary participation of crowd labor~\cite{Fang-participation}.

Many studies have focused on analyzing individuals' motivations and the incentives for participating in OSS projects~\cite{David-motivation-2008, Hemetsberger-motivation-2004, Lakhani-motivation-2005, Schofield-motivation-2006, Xu-motivation-2009, Ghosh-motivation-2005}.
Von Krogh et al.~\cite{Krogh-motivation-2012} classified contributors' motivations into three categories, namely, intrinsic motivation (\emph{e.g.}, ideology and fun), internalized extrinsic motivation (\emph{e.g.}, reputation and own use), and extrinsic motivation (\emph{e.g.}, career and pay).
Among developers who volunteer to contribute to open source projects, their motivation is mainly intrinsic or internalized extrinsic motivation~\cite{Krogh-motivation-2012}. They have full-time jobs and spend some spare time making open source contributions~\cite{open-source-get-paid}. However, Hars et al.~\cite{Hars-motivation} found that being paid can promote continuous contribution from developers with all types of motivation.

Currently, there are many ways to obtain financial support for open source initiatives, \emph{e.g.}, through donations or bounties.
Many studies have focused on the characteristics, impact, or effectiveness of each form of financial support.
For example, regarding bounties, Zhou et al.~\cite{Zhou-bountysource} studied the relation between issue resolution and bounty usage and found that adding bounties would increase the likelihood of issue resolution. Acting as a way for recruiting developers, setting bounties attracts those developers who want to make money through open source contributions, which facilitate the completion of complex tasks.
However, unlike bounty, the donation is a way of passively obtaining financial support.
Regarding open source donation,
Krishnamurthy et al.~\cite{Krishnamurthy-donation} studied the donation to the OSS platform and found the relation between donation level and platform association length and relational commitment.
For the donation to OSS, 
Nakasai et al.~\cite{Nakasai-donation, Nakasai-donation-eclipse-analysis} analyzed the incentives of individual donors and found that the benefits for donors and software release could promote donations. In contrast, bugs in software will negatively affect the number of donations. However, they only focused on eclipse projects.
Overney et al.~\cite{Overney-rich} studied the impact of donations from a broader perspective of open source projects on GitHub, which corresponds to NPM packages and explicitly mentions the way of donation in the README.md files. They found that only a small fraction (mainly active projects) asked for donations, and the number of received donations was mainly associated with project age. Most donations are requested and eventually used for engineering activities. However, there was a slight influence of donation on project activities.
Although Overney et al. did a thorough analysis of project-level donation, there lacks analysis of donation towards open source developers. Also, we think adding the qualitative analysis from the users' perspective can confirm the quantitative findings and help understand the pros and cons of system design and use.

\section{Background}
\label{background}

\subsection{Terminology}
\label{terminology}
To help the reader understand the rest of the article, we introduce key terms related to the \textsc{Sponsor} mechanism.

\emph{Sponsor}:
\textbf{an entity who provides donations to others}.

\emph{Maintainer}: \textbf{an entity who can be sponsored (developers who set up a \textsc{Sponsor} profile)}.

\emph{Nonmaintainer}: \textbf{an entity who has not set up the \emph{Sponsors}}.

\emph{Sponsorship}: \textbf{the donation relationship between a \emph{sponsor} and a \emph{maintainer}}.

\emph{AccountSetUpTime}: \textbf{the time when \emph{maintainers} set up the \textsc{Sponsor} profile for their accounts}.

\emph{FirstSponsorTime}: \textbf{the time when \emph{maintainers} receive their first \emph{sponsorship}}.

\subsection{Introduction of the \textsc{Sponsor} mechanism}
\label{sponsor-introduction}

\begin{figure}[htbp]
	\centering
	\includegraphics[width=\linewidth]{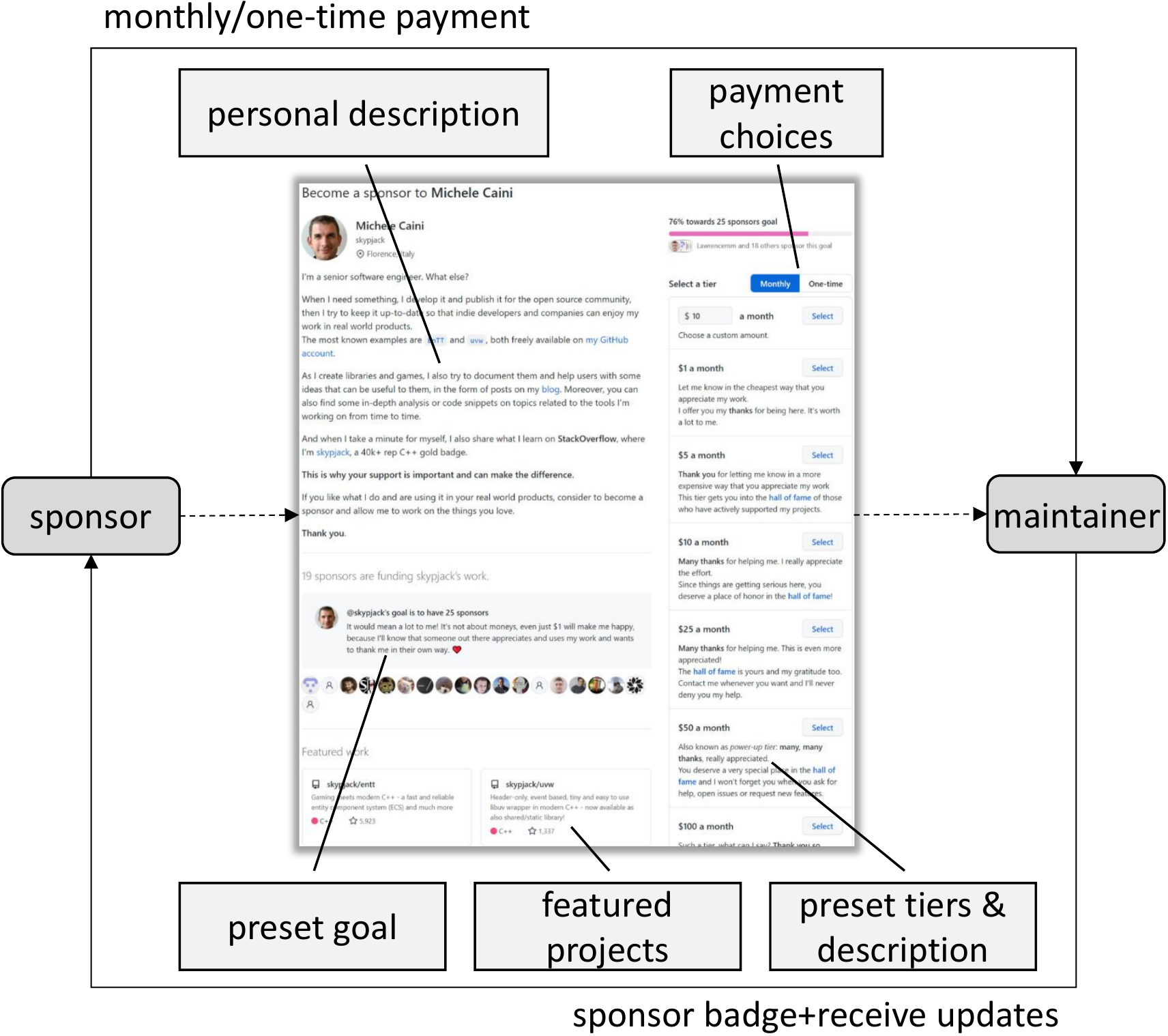}
	\caption{The workflow and key elements of \emph{sponsorship}}
	\label{fig:sponsorflow}
	\Description{This figure introduces the workflow of establishing \emph{sponsorship}, which also includes the related elements in the current GitHub's \textsc{Sponsor} mechanism.}
\end{figure}

Currently, in GitHub, the workflow and key elements of \emph{sponsorship} are shown in Figure~\ref{fig:sponsorflow}, where the \emph{sponsorship} is constructed on the \emph{maintainer}'s sponsor page by clicking the "select" button of specific amount.
The sponsor page is preset by the \emph{maintainer} when setting up a \textsc{Sponsor} profile in the related GitHub account, which mainly consists of the following elements.
\begin{itemize}
	\item Personal description: \emph{maintainers} are free to add text and modify it at any time. The main content can cover basic personal information, project information, why they need to be sponsored, other ways of donation, etc.
	\item Preset goal: \emph{maintainers} are allowed to set the number of \emph{sponsors} or \emph{sponsorships} that they want to get from the \textsc{Sponsor} mechanism and add related descriptions about the goal.
	\item Featured projects: this part lists the related projects that the \emph{maintainer} currently works on or with the most popularity.
	\item Preset tiers \& description: this part contains the tiers set by the \emph{maintainer}. \emph{Sponsors} can choose which tier to pay according to the amount and the related description.
	\item Payment choices: \emph{sponsors} can choose to monthly or one-time customized payment.
\end{itemize}
After choosing the way to construct the \emph{sponsorship}, \emph{sponsors} can get the sponsor badge and receive updates from the sponsored \emph{maintainer} in the future.

\subsection{Preliminary analysis}

\begin{figure}[htbp]
	\centering
	\includegraphics[width=\linewidth]{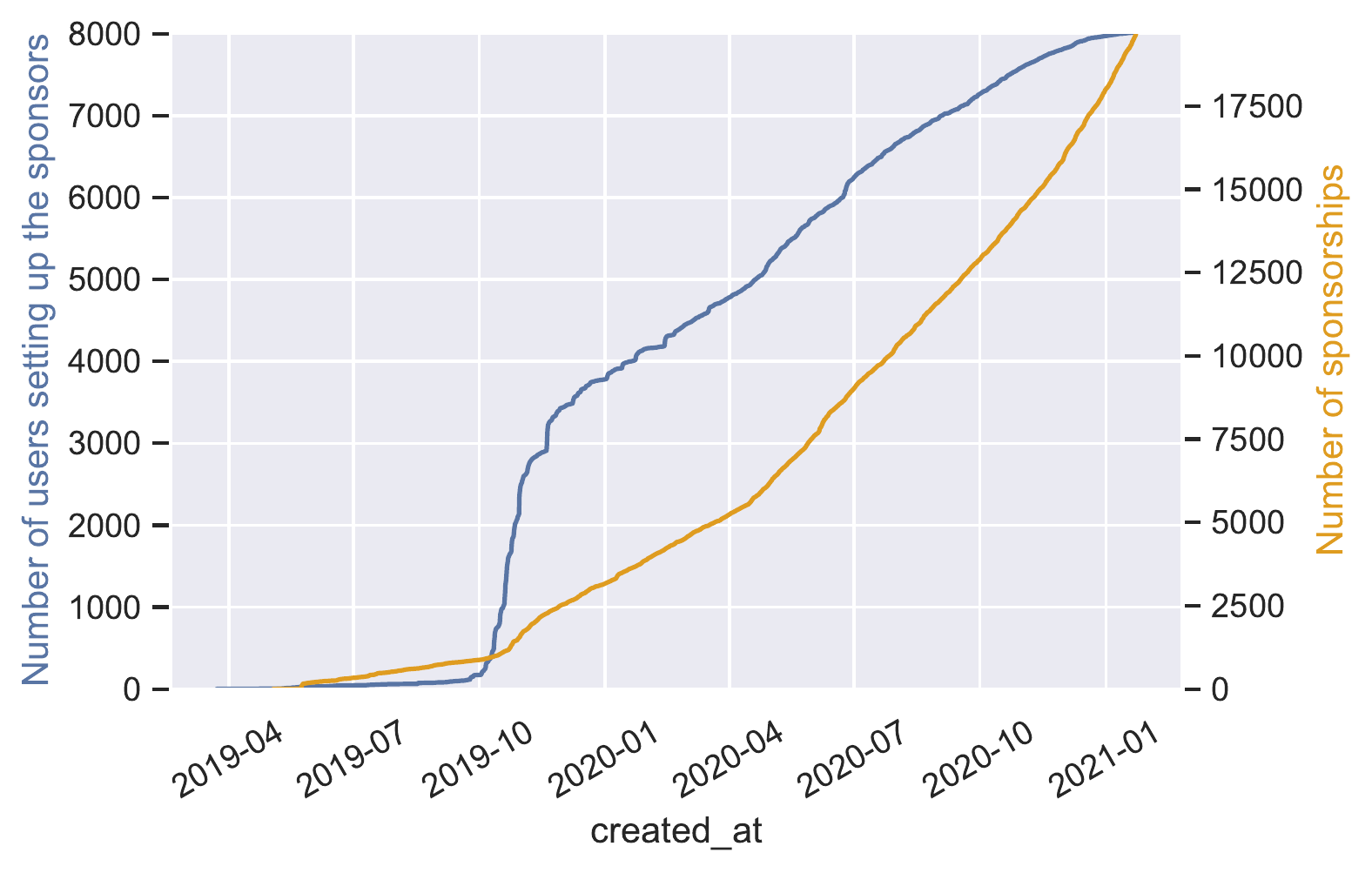}
	\caption{Cumulative participation over time}
	\label{fig:change-cumulative}
	\Description{This figure presents two changes: 1. how the number of users who set up the \textsc{Sponsor} account increase from 2019-05 to 2021-01. It increases slowly after 2019-10. 2. how the number of created sponsorship increase from 2019-05 to 2021-01. It is almost linear with a slight upward trend.}
\end{figure}

We conduct a statistical analysis of the use trends of the \textsc{Sponsor} mechanism
(Figure~\ref{fig:change-cumulative} shows the number of developers who set up the \textsc{Sponsor} account and how the number of sponsorships changes over time).
We can see that the number of developers who set up an account increased sharply around October 2019 (new things inspire people's interest). At other times, the growth rate shows a downward trend.
Meanwhile, the absolute number of participants in this mechanism increased steadily, although the growth rate shows a slight upward trend. Compared to GitHub itself, which has shown a strong increase in its user base \cite{Yu-exploring}, the \textsc{Sponsor} mechanism has not attracted as much attention.
In this context, we formulate \textbf{RQ1: Why do individuals participate (or not) in the \textsc{Sponsor} mechanism?}

According to our manual observation of GitHub developers' sponsorship pages, we find that developers can spend more time on their open source work if sponsored by others (with examples of this trend being Tim Condon \cite{github-Tim-Condon} and Super Diana \cite{github-Super-Diana}).
In short, we consider how the \textsc{Sponsor} mechanism may affect developers' open source activities. In this context, we ask \textbf{RQ2: How effective is sponsorship in motivating developer OSS activity?}

\begin{figure}[htbp]
        \centering
        \includegraphics[width=\linewidth]{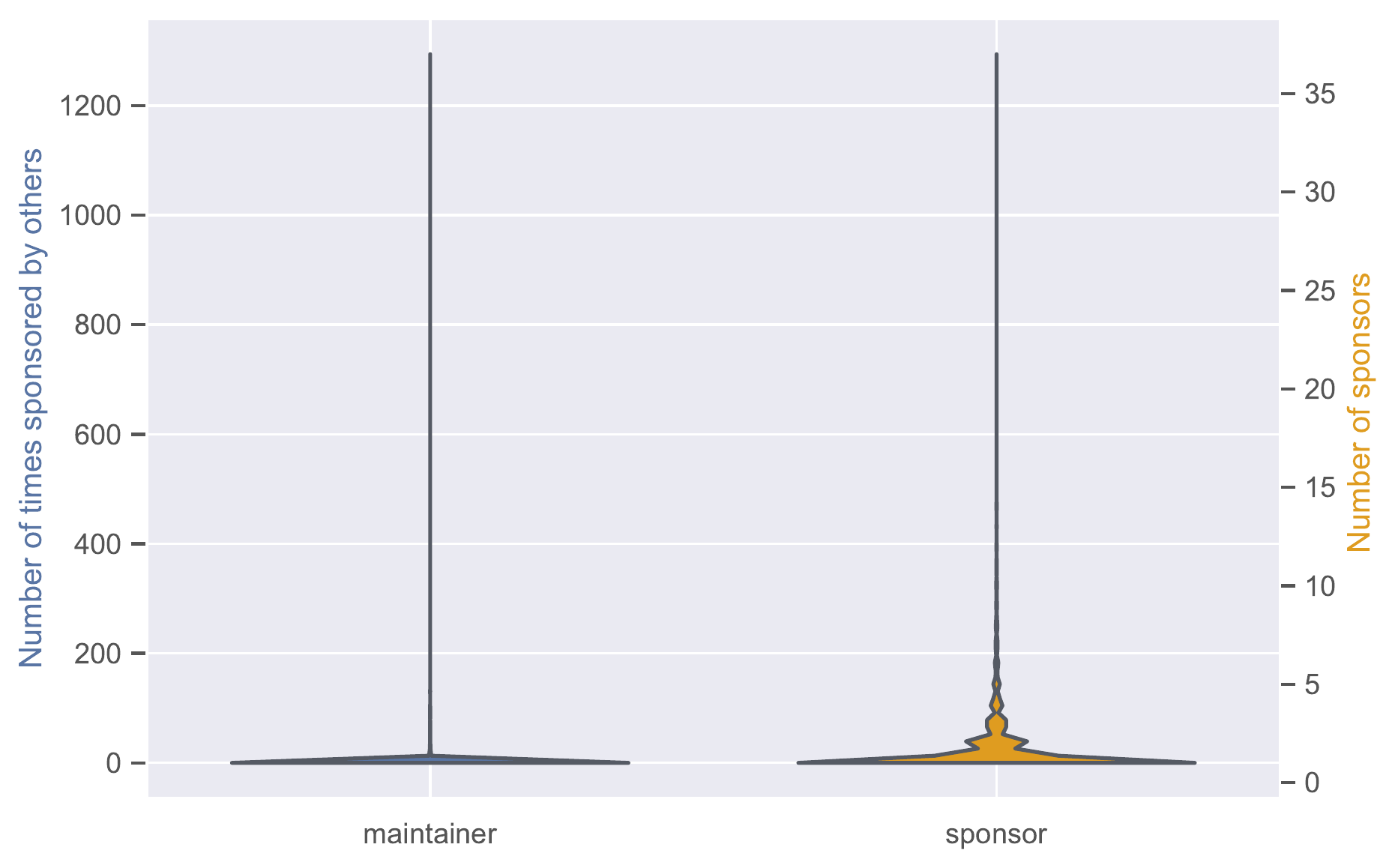}
        \caption{Differences between users in participation}
        \label{fig:user-difference-times}
        \Description{This figure presents the number of times maintainers are sponsored by others and the number of sponsors for sponsors. The two violin plots have two different axes, where the left is for the maintainers and the right is for the sponsors.}
\end{figure}

There are some very successful cases of individuals receiving support under the GitHub \textsc{Sponsor} mechanism (\emph{e.g.,} Caleb Porzio, who was sponsored by 1,314 sponsors as of 7 August 2021~\footnote{https://github.com/sponsors/calebporzio}).
However, most \textsc{Sponsor} participants have not been successful, and many have not received any sponsorships at all.
According to Figure~\ref{fig:user-difference-times}, only 14.1\% of \emph{maintainers} are sponsored at least once. Most people do not receive any sponsorships, despite setting up a \textsc{Sponsor} account.
Among \emph{sponsors}, most (76.3\%) sponsor others just one time.
Based on the statistical analysis results, we consider which developer characteristics lead to more sponsorships. In this vein, we ask \textbf{RQ3: Who is likely to receive more sponsorships?}

Currently, there are many ways to obtain financial support for open source initiatives, \emph{e.g.}, through donations or bounties.
The different types of financial support each have advantages and disadvantages~\cite{nayafia-financial}.
It falls to participants (especially those who have participated in multiple financial support mechanisms) to judge the reasonableness and effectiveness of each. 
To better understand users' perceptions of the \textsc{Sponsor} mechanism and thus enrich and improve it, we propose \textbf{RQ4: What are the shortcomings of the \textsc{Sponsor} mechanism?}

\section{Study Overview}
\label{method}

\subsection{Overall research methodology}
\begin{figure*}[htbp]
  \centering
  \includegraphics[width=0.8\linewidth]{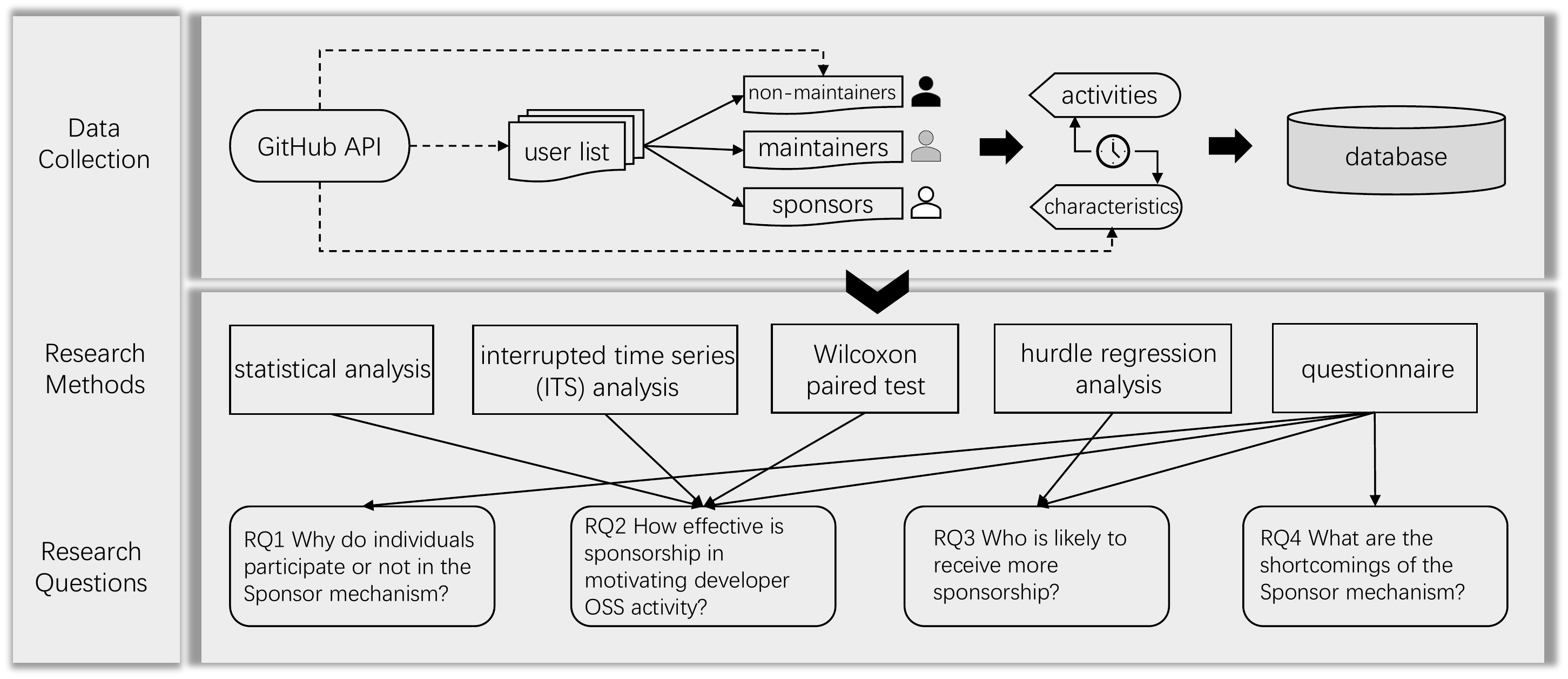}
\caption{Framework of this paper}
  \label{fig:framework}
  \Description{This is the framework of this paper, which contains 3 parts from the top to the bottom: data collection, research methods, and research questions. It describes how we get the data and how research methods are related to each research question.}
\end{figure*}

The overall framework of this paper is shown in Figure \ref{fig:framework}, with the research methodology consisting of two main parts: data collection and research methods.

\subsubsection{Data collection}

The data is collected using GitHub API.
The goal was to find different kinds of GitHub users (\emph{maintainers}, \emph{sponsors}, and \emph{nonmaintainers}) and gather their related basic information and activities.
Here, we focus on how to distinguish different kinds of users. The acquisition of relevant basic information and details on activities is described in the subsequent section (see Section \ref{detailed-research-methods}) when we introduce each research method in detail.
We acquired different types of users through the following steps.
\begin{enumerate}
\item We used the RESTful API \cite{github-api-list-users} to obtain all users. After that, we queried \emph{maintainers}
using the field \texttt{hasSponsorsListing} of the GraphQL API~\cite{github-objects-user}. We obtained 60,732,250 users who had not deleted their accounts, among which 7,992 users were individual \emph{maintainers}.
\item We used the field \texttt{sponsorshipsAsMaintainer} of the GraphQL API \cite{github-objects-user} to look up all the sponsorships that \emph{maintainers} had received and the corresponding \emph{sponsors}.
\item Using the list of \emph{sponsors} queried in step (2), we used the field \texttt{sponsorshipsAsSponsor} of the GraphQL API \cite{github-objects-user} to query all the related \emph{maintainers}. This step was to supplement the information on the \emph{maintainers} who had set up the \textsc{Sponsor} profiles identified during the query process in step (1).
\item We repeated steps (2) and (3) until no new \emph{maintainers} or \emph{sponsors} appeared.
\end{enumerate}
Through the above steps, we obtained 20,579 users, among which 8,028 are \emph{maintainers}, 13,555 are \emph{sponsors} (1,004 users are maintainers while sponsoring others at the same time). We also get 22,515 times of sponsorships. All users except maintainers were marked as \emph{nonmaintainers}.

\subsubsection{Research methods}
To answer the research questions, we used a combination of quantitative and qualitative analysis.
Regarding our \textbf{why} (\textbf{RQ1}) and \textbf{what} (\textbf{RQ4}) questions, since it was difficult to capture everyone's reasons for participation or nonparticipation and summarize the shortcomings of the mechanism based on just the platform information, we asked relevant people to complete a questionnaire.
For the \textbf{how} (\textbf{RQ2}) and \textbf{who} (\textbf{RQ3}) questions, we collected maintainer-related data, quantitatively analyzed the impact of sponsorship behavior on \emph{maintainer} open source activity, and explored the correlation between factors and the amount of sponsorship.
On this basis, we again conducted a qualitative analysis using a questionnaire.
This combination of quantitative and qualitative analysis led to our conclusions.
Next, we describe each research method in detail.

\subsection{Detailed introduction of research methods}
\label{detailed-research-methods}

\subsubsection{Questionnaire}
Since there are three types of interaction between the user and the \textsc{Sponsor} mechanism, namely, interactions with a \emph{sponsor}, a \emph{maintainer}, or a \emph{nonmaintainer} (see Section \ref{terminology}),
we designed three different online surveys \cite{questionnaire-design}.
The surveys for both \emph{sponsors} and \emph{maintainers} relate to their expectations for and satisfaction with the \textsc{Sponsor} mechanism. The survey for \emph{nonmaintainers} relates to their reason for not setting up the \textsc{Sponsor} feature for their account.
All the surveys start with an introduction to the research background and purpose.
There are two types of questions in each survey.
\begin{itemize}
\item Demographic questions designed to obtain participants' information, including their role in and experience with OSS development (the predefined answers were inspired by prior research \cite{Li-abandonment}).
\item Main questions, designed to gather users' views on the \textsc{Sponsor} mechanism.
\end{itemize}
Among the main questions, there are three kinds.
\begin{itemize}
\item Open-ended questions aimed at gathering answers.
\item Rating scale questions soliciting users' satisfaction and agreement levels.
\item Multiple-choice questions with ``Other'' text field options aimed at gathering large-scale user feedback while providing additional answers.
\end{itemize}
We provide a final, open-ended question to allow participants to talk freely about the \textsc{Sponsor} mechanism.
We discussed the questions with software engineering researchers to ensure that the items were well designed for our study and clear enough for participants to answer.
Finally, we used SurveyMonkey \cite{surveymonkey} to deploy our online surveys.

There were two rounds of each survey: 1) the \emph{pilot stage}, aimed at gathering answers to the open-ended questions from a limited number of participants, and 2) the \emph{full-scale stage}, aimed at gathering the votes for each answer from a larger population.
The statistics on the two stages can be seen in Table~\ref{statistics-questionnaire-recruitment}.

\begin{table}[htbp] \centering
    \setlength\tabcolsep{2.5pt}
    \renewcommand\arraystretch{1.00}
    \footnotesize
    \caption{Statistics on the two-stage survey}
    \label{statistics-questionnaire-recruitment} 
    \begin{tabular}{clccc}
    \toprule
    \textbf{Stage} & \textbf{Statistic items} & \textbf{Maintainers} & \textbf{Sponsors} & \textbf{Nonmaintainers} \\
     \cmidrule(r){1-2} \cmidrule(r){3-3} \cmidrule(r){4-4} \cmidrule(r){5-5}
    \multirow{4}{*}{\emph{Pilot}} & \#selected participants & 400 & 400 & 400 \\
      & \#successful invitations & 394 & 388 & 390 \\
      & \#response (\%) & 45 (11.4\%) & 24 (6.2\%) & 9 (2.3\%) \\
      & Date for collection & \multicolumn{3}{c}{June 8, 2021 - June 15, 2021} \\
    \cmidrule(r){1-5}
    \multirow{4}{*}{\emph{Full-scale}} & \#selected participants & 6,104 & 6,359 & 7,500 \\
    & \#successful invitations & 5,951 & 6,224 & 7,343 \\
    & \#response (\%) & 467 (7.8\%) & 396 (6.4\%) & 202 (2.8\%) \\
    & Date for collection & \multicolumn{3}{c}{June 29, 2021 - July 13, 2021} \\
    \bottomrule
    \multicolumn{5}{l}{\# means the number, \emph{e.g.,} \#response implies the number of responses} \\
    \end{tabular}
\end{table}

\paragraph{Participant recruitment.}
To recruit participants for the two rounds of three different surveys, we took the following steps:
\begin{enumerate}

\item For all three types of users (\emph{maintainers}, \emph{sponsors}, \emph{nonmaintainers}), we filtered out those whose email or name information could not be openly accessed, as these users might not want to receive questionnaires.

\item For all three types of users, we filtered out those who had not been active in the last month (since May 3, 2021), as they might not have focused on open source work on GitHub in recent days. In this step, we used the GitHub API to obtain users' recent activity, including the top repositories to which they had contributed in the last month and their last update time (field "updatedAt") on GitHub \cite{github-objects-user}.

\item For \emph{nonmaintainers}, we selected only users who may be eligible to set up a \textsc{Sponsor} profile based on their location information and the list of countries or regions included under the GitHub \textsc{Sponsor} mechanism \cite{github-objects-sponsor}.

\item After completing the above three steps, we randomly selected 400 unique individuals of each type without overlap as participants in the \emph{pilot stage}.

\item For the \emph{full-scale stage}, we selected all other \emph{maintainers} (6,104) and \emph{sponsors} (6,359) as participants. For \emph{nonmaintainers}, due to the low response rate in the \emph{pilot stage}, we filtered users according to the total number of stars of projects owned by developers (collected on 23 June 2021). We selected those with at least ten stars (we assumed that developers with popular projects are more likely to be interested in the \textsc{Sponsor} mechanism and use GitHub very often). After that, we randomly selected 7,500 participants.
\end{enumerate}

\paragraph{Response and analysis.}

After selecting the participants, we published the questionnaire online and sent the web address to participants via email. The email invitation contained the basic information of the questionnaire publisher, the reason for the release, the number of questions, and the estimated time required to fill out the questionnaire.


Based on the participants' feedback of the \emph{pilot} stage, we designed the questionnaires for the \emph{full-scale stage}. We removed 1 question for \emph{maintainers}, 1 question for \emph{sponsors}, and 2 questions for \emph{nonmaintainers} due to answers with repetitive content in relation to the answers to other questions. We extracted the essential information from all responses and turned some open questions into multiple-choice questions (3 for \emph{maintainers}, 3 for \emph{sponsors}, and 1 for \emph{nonmaintainers}) through open coding of card sorting method~\cite{Zimmermann-card} by the first, second and the fifth authors together.
To avoid disturbing the participants, we extended the time to collect the responses in this stage relative to that in the \emph{pilot} stage but did not send a second email reminder. At the same time, because different types of participants dedicate different amounts of attention to the \textsc{Sponsor} mechanism, the response rate varies greatly. \emph{Nonmaintainers}, who do not participate in the \textsc{Sponsor} mechanism, may not care about it and not want to reply to the email.

When analyzing the multiple-choice questions, we first calculated the voting rate for each preset option. After that, we manually included the textual response for the ``Other'' option into the preset taxonomy, if possible, via the closed coding method \cite{Zimmermann-card}. If a new topic emerged, we integrated it into the existing taxonomy.
When analyzing the last open question (``Do you have anything else to tell us about the \textsc{Sponsor} mechanism?''), we extracted the essential information from the textual response for qualitative analysis.
To facilitate analysis, we use \emph{[MCx]}, \emph{[SCx]}, and \emph{[OCx]} to represent the textual response in the questionnaire for \emph{maintainers}, \emph{sponsors}, and \emph{nonmaintainers}, respectively, where \emph{x} indicates the serial number of the comment.

Through the first two questions of each questionnaire, we collected participants' demographic information, including their status and experience with open source development.
For the \emph{full-scale stage}, the results are shown in Table \ref{result-demographic}.
More than 70\% of participants in each category have more than three years of OSS development experience. More than 10\% of \emph{sponsors} have no OSS development experience, which indicates that many \emph{sponsors} sponsor others solely to support OSS development or maintenance.

\begin{table}[htbp] \centering
  \setlength\tabcolsep{1pt}
  \renewcommand\arraystretch{1.00}
  \scriptsize
  \caption{Demographic information of participants in the \emph{full-scale stage}}
  \label{result-demographic} 
  \begin{tabular}{p{0.3\linewidth} p{0.35\linewidth} p{0.1\linewidth} p{0.1\linewidth} p{0.1\linewidth}}
  \toprule
  \textbf{Questions} & \textbf{Answers} & \textbf{M (\%)} & \textbf{S (\%)} & \textbf{NM (\%)} \\
  \cmidrule(r){1-1}  \cmidrule(r){2-2} \cmidrule(r){3-3} \cmidrule(r){4-4} \cmidrule(r){5-5}
  \multirow{4}{*}{\shortstack{Q1: How would you best\\ describe yourself?}} & Developer working in industry & 62.3 & 80.0 & 65.5 \\
    & Full time independent developer & 16.6 & 10.0 & 8.0 \\
    & Student & 11.6 & 6.9 & 6.5 \\
    & Academic researcher & 3.7 & 3.6 & 16.0 \\
  \cmidrule(r){1-5}
  \multirow{5}{*}{\shortstack{Q2: How many years of OSS\\ development experience\\ do you have?}} & Never & 1.1 & 10.2 & 3.0 \\
    & <1 year & 2.2 & 4.6 & 6.5 \\
    & 1-3 years & 10.1 & 14.5 & 12.6 \\
    & 3-5 years & 21.9 & 22.6 & 23.1 \\
    & 5-10 years & 33.6 & 26.9 & 27.1 \\
    & >10 years & 31.2 & 21.3 & 27.6 \\
  \bottomrule
  \multicolumn{5}{l}{M: \emph{maintainer}; S: \emph{sponsor}; NM: \emph{nonmaintainer}}\\
  \end{tabular}
\end{table}


\subsubsection{ITS analysis}
\label{method-ITS}

The aim of this analysis was to determine when to treat sponsorship as an \emph{intervention} and how it influences the potential trends in \emph{maintainers}' activities (development and discussion activities) from a long-term perspective. Therefore, following the guidelines of previous studies \cite{Overney-rich,Zhao-impact,Trockman-sparkle}, we used the ITS method.
The settings of the ITS analysis are shown below.

\emph{Interventions}: We set both \emph{accountSetUpTime} and \emph{firstSponsorTime} (see Section \ref{terminology}) as separate interventions. We assumed that \emph{maintainers} may increase their activity after \emph{accountSetUpTime} to attract others' attention for future sponsorship or be motivated to increase their open source contributions after \emph{firstSponsorTime}.

\emph{Responses}: We set the \emph{number of commits} (development activity) and the \emph{number of discussions} (discussion activity) as responses, as they indicate different kinds of activities on GitHub.

\emph{Unstable period}: Similar to previous studies \cite{Overney-rich,Zhao-impact,Trockman-sparkle}, we set 15 days before and after interventions as the unstable period.

\emph{Before \& after intervention periods}: To retain enough analyzable data, we selected \emph{maintainers} with at least six months of activity before and after interventions in addition to the unstable period. Therefore, each \emph{maintainer} has at least $15*2+6*2*30=390$ days of activity on GitHub.

\emph{Time window}: Each month in \emph{before \& after intervention periods} is a time window, and the unstable period is also a time window. Therefore, there are $6*12+1=13$ time windows in all.

The independent variables are as follows.

\paragraph{Basic items}
\begin{itemize}
\item \emph{intervention}: Binary variable indicating an intervention
\item \emph{time}: Continuous variable indicating the time by month from the start of an observation to each time window, with a value range of $[0, 12]$
\item \emph{time after intervention}: Continuous variable indicating how many months have passed after an intervention (if there is no intervention, \emph{time after intervention=0}; otherwise, \emph{time after intervention=time-6}).
\end{itemize}

\paragraph{Developer characteristics}
\begin{itemize}
\item \emph{number of stars before}: Continuous variable, measured as the total number of stars of \emph{maintainer}-owned repositories before the start of each time window
\item \emph{in company}: Binary variable indicating whether company information exists at data collection time
\item \emph{has goal}: Binary variable indicating whether a \emph{maintainer} sets a goal for sponsorship at data collection time
\item \emph{has another way}: Binary variable indicating whether a \emph{maintainer} sets other methods for receiving donations at data collection time
\item \emph{is hireable}: Binary variable indicating whether a \emph{maintainer} declares a hireable status at data collection time
\end{itemize}

\paragraph{Developer activities}
\begin{itemize}
\item \emph{number of commits before}: Continuous variable measured as the number of commits before the start of each time window
\item \emph{number of discussions before}: Continuous variable measured as the number of discussions before the start of each time window
\end{itemize}

We built a mixed effect linear regression model for ITS analysis with a \emph{maintainer} identifier as the random effect and all the measured factors as fixed effects.
A major advantage of the mixed effect model is that it can eliminate the correlated observations within a subject \cite{linear_mixed_effect}. Here, the time windows for the same \emph{maintainer} tend to have a similar trend.
We used the \emph{lmer} function of the \emph{lmerTest} package in R \cite{lmerTest-paper} to fit models for the \emph{maintainer's} commit and discussion activities.
For better model performance, we transformed the continuous variables to make them approximately normal and on a comparable scale through log-transformation (plus 0.5) and standardization (mean 0, standard deviation 1) \cite{feature-transformation}.
To reduce the multicollinearity problem, we excluded factors with variance inflation factor (VIF) values $\geq 5$ using the \emph{vif} function of the \emph{car} package in R \cite{cohen_applied}.
We report the coefficients and the related \emph{p} values obtained in this way.
We also report the explained variance of the factor, which can be interpreted as the effect size relative to the total variance explained by all the factors.
For the fitness of models, we report both marginal ($R^{2}_m$) and conditional ($R^{2}_c$) R-squared values using the \emph{r.squaredGLMM} function of the \emph{MuMIn} package in R \cite{MuMIn}.

Together with ITS analysis, we visually present how \emph{responses} change over time to show the activity change more intuitively (statistical analysis).
Since there is an \emph{unstable period} in the ITS analysis, we analyze this period separately using the \emph{Wilcoxon paired test} method, which is presented in the following section.

\subsubsection{Wilcoxon paired test}

For the ITS analysis, the unstable period is ignored.
However, the \textsc{Sponsor} mechanism involves a small amount of money, which may influence \emph{maintainer} behavior in the short term only. We assume that \emph{maintainers} may have great fluctuations in OSS activity during the unstable period.
We used a paired, nonparametric test method called the Wilcoxon paired test \cite{Canfora-wilcoxon}. Through two-sided tests (both \emph{alternative=greater} and \emph{alternative=less}) \cite{scipy-wilcoxon}, we can see whether the intervention increases or decreases a \emph{maintainer's} activity. We considered three kinds of interventions, including \emph{accountSetUpTime}, \emph{firstSponsorTime}, and before and after each sponsorship.
We used Cliff's delta ($\delta$) to measure the effect size \cite{Grissom-effect-size}, with $|\delta|<0.147$ indicating a negligible effect size, $0.147 \leq |\delta| < 0.33$ indicating a small effect size, $0.33 \leq |\delta| < 0.474$ indicating a medium effect size, and $|\delta| \geq 0.474$ indicating a large effect size.

\subsubsection{Hurdle regression analysis}

The critical idea of hurdle regression is to create a dataset with \emph{maintainer} characteristics and the amount of sponsorship established.
Therefore, we collected different characteristics of each \emph{maintainer} heuristically, including \emph{basic information}, \emph{social characteristics}, \emph{\textsc{Sponsor} mechanism characteristics}, \emph{developer activities}, and \emph{project characteristics}.
For the amount of sponsorship, we used the number of times that a \emph{maintainer} is sponsored.
Next, we present detailed descriptions of the collected variables.

\paragraph{Developer basic information}
\begin{itemize}
\item \emph{user age}: Continuous variable measured as the time interval by month since the creation of the user account in the GitHub community until the data collection time
\item \emph{in company}: Binary variable indicating whether a \emph{maintainer} introduces the personal work situation in detail
\item \emph{has email}: Binary variable indicating whether a \emph{maintainer} publicly provides the contact information
\item \emph{has location}: Binary variable indicating whether the \emph{maintainer} discloses the geographical location information
\item \emph{is hireable}: Binary variable indicating whether a \emph{maintainer} indicates availability for hire
\end{itemize}

\paragraph{Social characteristics}
\begin{itemize}
\item \emph{followers}: Continuous variable measured as the number of followers
\item \emph{followings}: Continuous variable indicating how many users the \emph{maintainer} follows
\end{itemize}

\paragraph{\textsc{Sponsor} mechanism characteristics}
\begin{itemize}
\item \emph{min tier}: Continuous variable measured as the minimum number of dollars set by the \emph{maintainer} for donations
\item \emph{max tier}: Continuous variable indicating the maximum donation
\item \emph{has goal}: Binary variable indicating whether a \emph{maintainer} sets a goal for sponsorship
\item \emph{has another way}: Binary variable indicating whether a \emph{maintainer} introduces other modes for receiving donations. Here, we identified other donation modes by finding links to other funding platforms in the description on the sponsorship page. Other platforms are shown in Table \ref{donation-platforms}, which was compiled according to the collection by Overney et al. \cite{Overney-rich} and the supported external links of GitHub \cite{github-funding-links}
\item \emph{introduction richness}: Continuous variable measured as the length of the introduction on the personal sponsorship page
\item \emph{user age after sponsor account}: Continuous variable indicating the time interval by month (to see how time influences the amount of sponsorship)
\end{itemize}

\paragraph{Activity characteristics}
\begin{itemize}
\item \emph{number of commits}: Continuous variable measured as the total number of commits in GitHub from \emph{accountSetUpTime} until the data collection time
\item \emph{number of discussions}: Continuous variable measured as the number of comments, including issue comments, pull request comments, and commit comments from \emph{accountSetUpTime} until the data collection time
\end{itemize}

\paragraph{Project characteristics}
\begin{itemize}
\item \emph{sum star number}: Continuous variable measured as the total number of stars of repositories created by a \emph{maintainer}
\item \emph{sum fork number}: Continuous variable indicating the number of forks
\item \emph{sum watch number}: Continuous variable indicating the number of watchers
\item \emph{sum top repository star number}: Continuous variable measured as the total number of stars of top repositories that a \emph{maintainer} contributed in the four months before data collection \cite{github-top-repo}
\item \emph{number of dependents}: Continuous variable measured as the number of repositories that rely on the project with the most watchers among all projects owned by the \emph{maintainer}
\end{itemize}

When building the hurdle regression models,
we removed \emph{maintainers} with less than 3 months of activity after \emph{accountSetUpTime} to reduce the impact of time on sponsorship.
We reasoned that \emph{sponsors} need time to find \emph{maintainers} to donate to.
To reduce the zero-inflation in the response variance, we used hurdle regression \cite{Zeileis-hurdle} by splitting the sample into two parts:
\begin{enumerate}
\item \emph{maintainers} who have not received any donations from others, to examine which factors influence whether a \emph{maintainer} receives donations
\item \emph{maintainers} with at least 1 sponsorship, to examine how the amount of received donations is influenced by the aforementioned characteristics
\end{enumerate}

For the reduction of the multicollinearity problem and the report of results, we use the same methods (see Section~\ref{method-ITS}).

\section{Results}
\label{results}

\subsection{RQ1: \textit{Why} do individuals participate or not in the \textsc{Sponsor} mechanism?}

For this research question, the questionnaire had a dedicated item for each of the three types of participants, \emph{i.e.,} \textbf{Q3} for \emph{maintainers}, \emph{sponsors}, and \emph{nonmaintainers}.
Table A shows the motivations or reasons elaborated by different types of developers in the \emph{full-scale stage} and the percentage of votes for each option.

\begin{table*}[htbp] \centering
    \setlength\tabcolsep{1pt}
    \renewcommand\arraystretch{1.00}
    \footnotesize
    \caption{Reasons for participating or not participating in the \textsc{Sponsor} mechanism}
    \label{result-reasons} 
  \begin{tabular}{p{4.3cm}<{\raggedright}p{0.4cm}<{\raggedleft}p{1cm}<{\raggedright}p{4.3cm}<{\raggedright}p{0.4cm}<{\raggedleft}p{1cm}<{\raggedright}p{4.3cm}<{\raggedright}p{0.4cm}<{\raggedleft}p{1cm}<{\raggedright}}
  \toprule
   \textbf{Reason\_maintainers} & \multicolumn{2}{l}{\textbf{Votes (\%)}} & \textbf{Reason\_sponsors} & \multicolumn{2}{l}{\textbf{Votes (\%)}} & \textbf{Reason\_non-maintainers} & \multicolumn{2}{l}{\textbf{Votes (\%)}} \\
   \cmidrule(r){1-3}  \cmidrule(r){4-6} \cmidrule(r){7-9}
   $\mathcal{RM}1$ It allows users of my projects to express thanks/appreciation & 64.9  & \includegraphics[width=6.49mm, height=2mm]{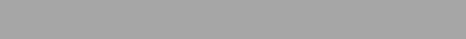} & $\mathcal{RS}1$ Because I benefit from the developer’s projects & 85.8  & \includegraphics[width=8.58mm, height=2mm]{pics/percentage.png} & $\mathcal{RO}1$ No need to be sponsored & 39.3  & \includegraphics[width=3.93mm, height=2mm]{pics/percentage.png} \\
    $\mathcal{RM}2$ Sponsorship can motivate my future OSS contribution & 63.1  & \includegraphics[width=6.31mm, height=2mm]{pics/percentage.png} & $\mathcal{RS}2$ To encourage the developer to continue the contribution & 78.4  & \includegraphics[width=7.84mm, height=2mm]{pics/percentage.png} & $\mathcal{RO}2$ I contribute to OSS not for money & 38.3  & \includegraphics[width=3.83mm, height=2mm]{pics/percentage.png} \\
    $\mathcal{RM}3$ Side income for OSS contribution & 60.6  & \includegraphics[width=6.06mm, height=2mm]{pics/percentage.png} & $\mathcal{RS}3$ To show my recognition of the developer’s work & 69.5  & \includegraphics[width=6.95mm, height=2mm]{pics/percentage.png} & $\mathcal{RO}3$ My work is not worth being sponsored & 28.4  & \includegraphics[width=2.84mm, height=2mm]{pics/percentage.png} \\
    $\mathcal{RM}4$ It can reflect community recognition for my work & 39.9  & \includegraphics[width=3.99mm, height=2mm]{pics/percentage.png} & $\mathcal{RS}4$ Because I’m interested in the developer’s projects & 49.0  & \includegraphics[width=4.9mm, height=2mm]{pics/percentage.png} & $\mathcal{RO}4$ Never heard of it & 26.4  & \includegraphics[width=2.64mm, height=2mm]{pics/percentage.png} \\
    $\mathcal{RM}5$ Just for fun & 28.9  & \includegraphics[width=2.89mm, height=2mm]{pics/percentage.png} & $\mathcal{RS}5$ To motivate the developer to work harder on a specific feature & 9.4   & \includegraphics[width=0.94mm, height=2mm]{pics/percentage.png} & $\mathcal{RO}5$ It’s cumbersome & 8.5   & \includegraphics[width=0.85mm, height=2mm]{pics/percentage.png} \\
    $\mathcal{RM}6$ I deserve to be rewarded for my past OSS contribution & 21.8  & \includegraphics[width=2.18mm, height=2mm]{pics/percentage.png} & $\mathcal{RS}6$ Because I know the developer & 8.9   & \includegraphics[width=0.89mm, height=2mm]{pics/percentage.png} & $\mathcal{RO}6$ Not available in my region & 2.0   & \includegraphics[width=0.2mm, height=2mm]{pics/percentage.png} \\
    $\mathcal{RM}7$ I am able to prioritize the requirements of sponsors (e.g., fixing bugs) & 18.8  & \includegraphics[width=1.88mm, height=2mm]{pics/percentage.png} & Other & 1.0   & \includegraphics[width=0.1mm, height=2mm]{pics/percentage.png} & Other & 10.4  & \includegraphics[width=1.04mm, height=2mm]{pics/percentage.png} \\
    $\mathcal{RM}8$ It’s a way for me to make a living & 13.1  & \includegraphics[width=1.31mm, height=2mm]{pics/percentage.png} &       &       &       &       &       &  \\
    Other & 1.9   & \includegraphics[width=0.19mm, height=2mm]{pics/percentage.png} &       &       &       &       &       &  \\

  \bottomrule
  \end{tabular}
  \end{table*}

\subsubsection{Related motivations}
From the results, we find that some of the motivations of \emph{maintainers} and \emph{sponsors} are related.

\emph{Project use relationship}. For \textbf{$\mathcal{RM}1$} and \textbf{$\mathcal{RS}1$}, they all indicate that the usage of related projects leads to sponsorship.
Some 64.9\% of \emph{maintainers} and 85.8\% of \emph{sponsors} cite this factor as one motivation for participating in the \textsc{Sponsor} mechanism; this consensus puts it in first place on both groups' motivation lists.
People think that users should give back to contributors in various ways, among which the \textsc{Sponsor} mechanism serves as a \emph{``nice way to say thanks''[MC23]} and \emph{``allow people to easily fund their projects.'' [MC20]}.
From the perspective of \emph{sponsors}, developers are grateful for the OSS that they use and hope to express their gratitude and, \emph{e.g.,} \emph{``show support for OSS, which I heavily rely on in my daily work. Without OSS, I could not have built a career in data science'' [SC3]}.

\emph{Promotion of continuous OSS contributions}. \textbf{$\mathcal{RM}2$} and \textbf{$\mathcal{RS}2$} reflect participants' uniform motivation to engage in further OSS contributions.
Some 63.1\% and 78.4\% of \emph{maintainers} and \emph{sponsors}, respectively, cite this factor as a motivation; this factor thus ranks 2nd among all the enumerated reasons for participation.
For open source developers, if they want to devote themselves to open source projects, they need to solve the problem of daily costs and open source maintenance costs (\emph{e.g., ``I believe in the open source and good-for-humanity idea. I need to get paid only to live a decent life'' [MC37]}).
Therefore, the emergence of the \textsc{Sponsor} mechanism may help them solve the above problems to a certain extent and then invest more time in open source projects
(\emph{e.g., ``I was really hoping to get sponsorship so I could spend more time focusing on developing open source projects'' [MC11]}).
For \emph{sponsors}, they also hope to inspire contributors to continue to make outstanding contributions
(\emph{e.g., ``motivate them to do the awesome work'' [SC5]}).

\emph{Recognition of OSS work}. For \textbf{$\mathcal{RM}4$} and \textbf{$\mathcal{RS}3$}, they all indicate \emph{sponsors'} recognition of \emph{maintainers}. A total of 39.9\% of \emph{maintainers} and 49\% of \emph{sponsors} cite this factor as a motivation for participation; this motivation ranks 4th and 3rd for these two groups, respectively.
For some people, sponsorship is a manifestation of greater recognition by \emph{sponsors} than income.

\emph{Support for specific features}. For \textbf{$\mathcal{RM}7$} and \textbf{$\mathcal{RS}5$}, 18.8\% of \emph{maintainers} and 9.4\% of \emph{sponsors} hope that the \textsc{Sponsor} mechanism can help set the agenda for issue resolution priorities, although many people think that OSS should not be related to money (\emph{e.g., ``If there was money given by others involved, I would feel pressed to implement whatever they want (like in industry projects). I want FLOSS to be completely independent from corporate requests'' [OC5]}).

\subsubsection{Motivation across different user types}
In addition to the motivations mentioned above related to the \emph{sponsor} and \emph{maintainer} relationship, there are other motivations or reasons related to the kinds of users.

\emph{Maintainers}:
More than 60\% of these participants chose \textbf{$\mathcal{RM}3$}, but only 13\% chose \textbf{$\mathcal{RM}8$}. In the \textbf{Other} option, 4 participants mentioned that they hope sponsorship can cover some of their infrastructure costs. Moreover, 28.9\% of participants even chose \textbf{$\mathcal{RM}5$ Just for fun}. This indicates that different people have different expectations for the income. The \textsc{Sponsor} mechanism, as a donation mode, can bring maintainers a small amount of side income, but it is hardly a means for making a living (\emph{e.g., ``Sponsor is not an earning medium but a support mechanism'' [MC91]}).

\emph{Sponsors}:
Some 49\% of these respondents chose \textbf{$\mathcal{RS}4$}. They may not have used the \emph{sponsors'} project but may be interested only in contributions or projects. We also find that only 8.9\% of these participants chose \textbf{$\mathcal{RS}6$}, which indicates that the \emph{sponsor} may not know the \emph{maintainer} before the sponsorship. Therefore, social relationships are not particularly relevant to donations.

\emph{Nonmaintainers}:
Among developers who have not set up the \textsc{Sponsor} feature in their user account, we find that 38.3\% chose \textbf{$\mathcal{RO}2$}. This is related to the results on users setting up their \textsc{Sponsor} profile. There have always been two perspectives in the open source spheres. Some people think that money is needed to maintain projects even if they are open source; others feel that open source projects should be free.
In Table \ref{result-demographic}, we can see that the proportion of full-time independent developers is lowest among \emph{nonmaintainers}. At the same time, we examined why full-time independent developers do not set up the \textsc{Sponsor} feature in their accounts. Only 3 out of these 16 developers chose \textbf{$\mathcal{RO}2$} (however, they all had successful products). This indicates that full-time independent developers recognize and hope to gain benefits through open source contributions, and the sponsorship mechanism can be a way to meet their needs.
There are other reasons for not participating in the \textsc{Sponsor} mechanism. Some 39.3\% of participants do not currently need to be sponsored (\textbf{$\mathcal{RO}1$}), 28.4\% play down the value of their own work (\textbf{$\mathcal{RO}3$}),
and 26.4\% report that they have never heard of it (\textbf{$\mathcal{RO}4$}).
Others complain about problems with the mechanism, including its cumbersomeness (8.5\% -- \textbf{$\mathcal{RO}5$}), unavailability (2\% -- \textbf{$\mathcal{RO}6$}), and tax problems (4\% -- one reason cited under the ``Other'' option).

\begin{tcolorbox}[left=0mm, right=0mm, top=0mm, bottom=0mm]
The main reason cited for participation is to obtain or express appreciation for the use of open source projects or to recognize the \emph{maintainer}'s OSS contribution.
In turn, such support 
may promote better contributions.
\emph{Maintainers} seeking to make money tend to obtain extra income rather than a full livelihood through sponsorship.
For \emph{nonmaintainers}, in addition to personal reasons, the mixing of open source projects and money is another critical consideration preventing them from participating.
\end{tcolorbox}

\subsection{RQ2: \textit{How} effective is sponsorship in motivating developer OSS activity?}

We used the following methods for this research question: statistical analysis (visualization), ITS analysis, unstable period analysis based on the Wilcoxon paired test method, and qualitative analysis based on a questionnaire survey. We also explored the two kinds of interventions, namely, \emph{accountSetUpTime} and \emph{firstSponsorTime}.

\subsubsection{Visualization}
\begin{figure*}[htbp]
\begin{minipage}[t]{0.24\textwidth}
      \centering
      \includegraphics[width=\textwidth]{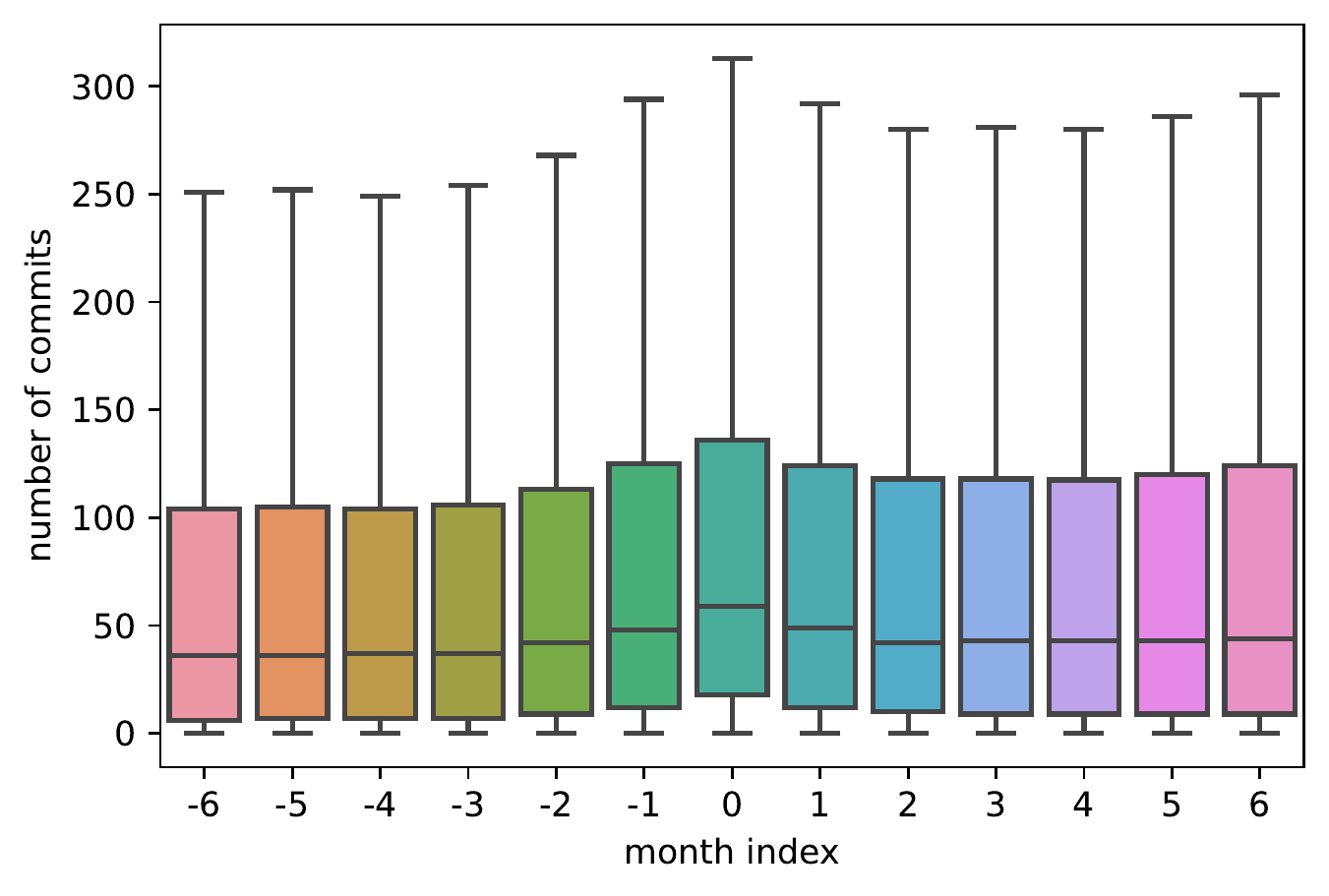}
\caption{\emph{Number of commits} before and after \emph{accountSetUpTime}}
      \label{fig:visual-commit-account}
      \Description{This figure shows the boxplots of the number of commits in each time window, where the median values of month index -1, 0, and 1 are higher than others, and months are split by the time when setting up the \textsc{Sponsor} account.}
\end{minipage}\hspace{1.5mm}
\begin{minipage}[t]{0.24\textwidth}
      \centering
      \includegraphics[width=\textwidth]{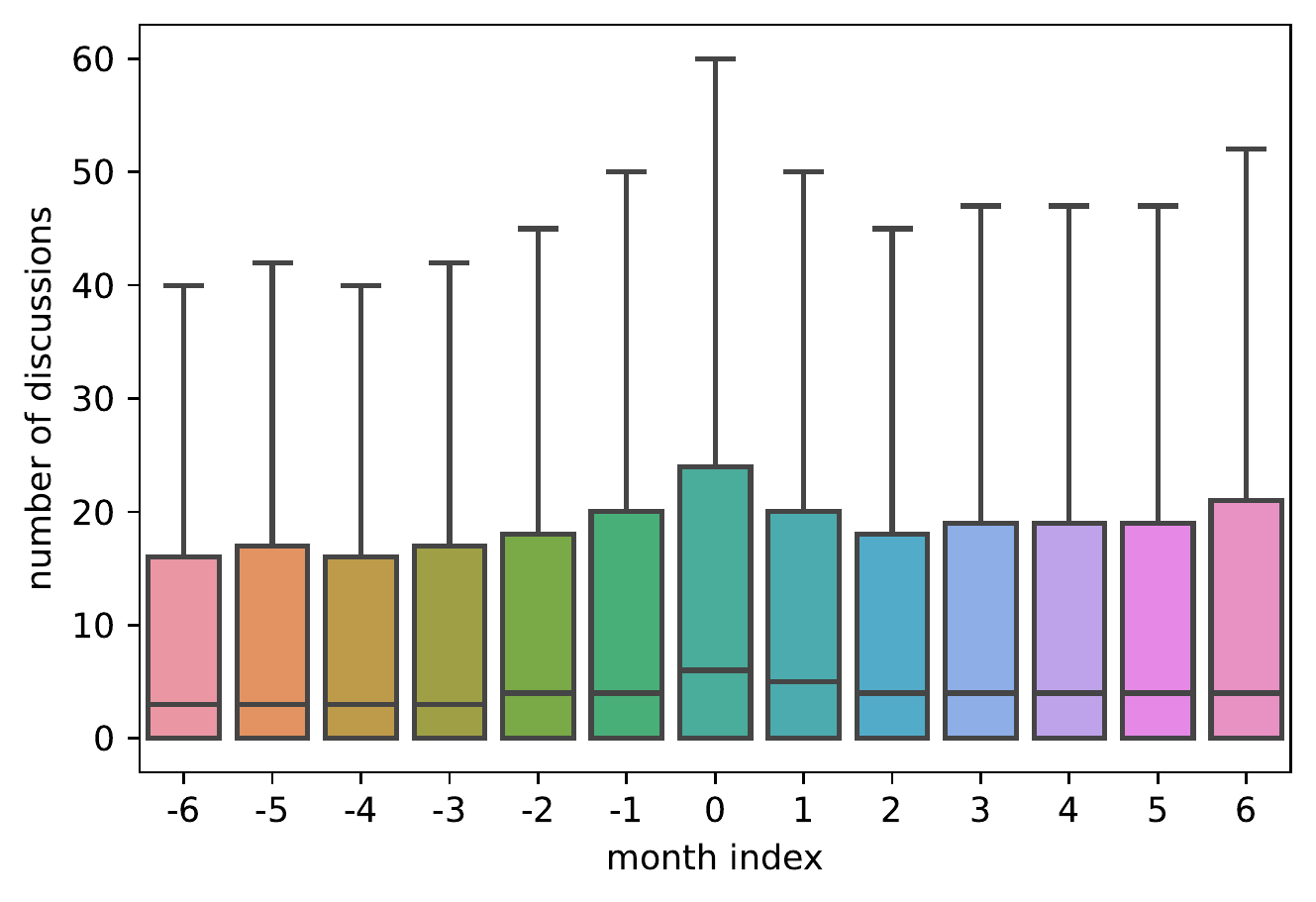}
\caption{\emph{Number of discussions} before and after \emph{accountSetUpTime}}
      \label{fig:visual-discussion-account}
      \Description{This figure shows the boxplots of the number of discussions in each time window, where the median values of month index -1, 0, and 1 are higher than others, and months are split by the time when setting up the Sponsor account.}
\end{minipage}\hspace{1.5mm}
\begin{minipage}[t]{0.24\textwidth}
    \centering
    \includegraphics[width=\textwidth]{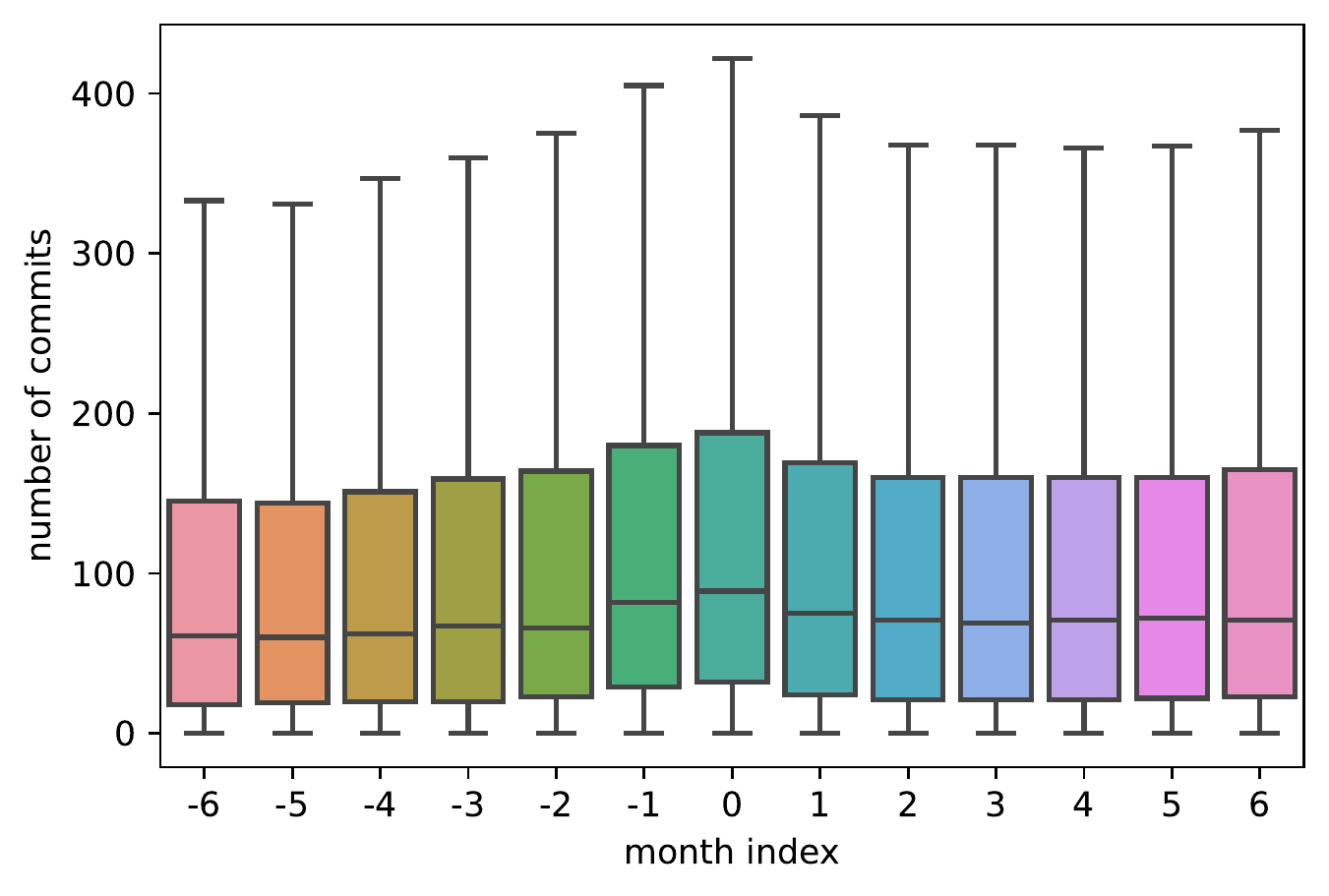}
\caption{\emph{Number of commits} before and after \emph{firstSponsorTime}}
    \label{fig:visual-commit-first}
    \Description{This figure shows the boxplots of the number of commits in each time window, where the median values of month index -1, 0, and 1 are higher than others, and months are split by the first time of receiving sponsorship.}
\end{minipage}\hspace{1.5mm}
\begin{minipage}[t]{0.24\textwidth}
      \centering
      \includegraphics[width=\textwidth]{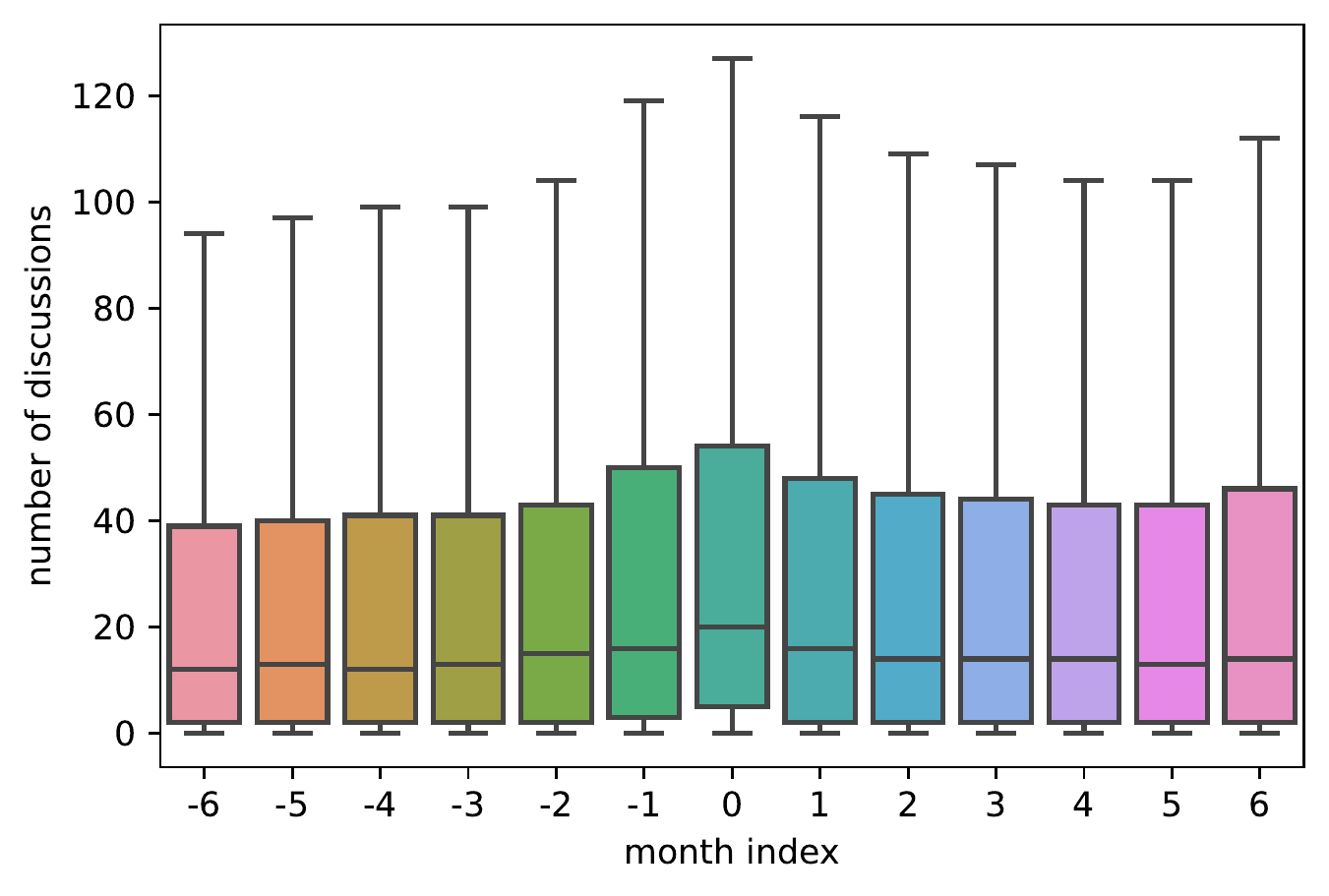}
\caption{\emph{Number of discussions} before and after \emph{firstSponsorTime}}
      \label{fig:visual-discussion-first}
      \Description{This figure shows the boxplots of the number of discussions in each time window, where the median values of month index -1, 0, and 1 are higher than others, and months are split by the first time of receiving sponsorship.}
\end{minipage}
\end{figure*}

Figures \ref{fig:visual-commit-account}-\ref{fig:visual-discussion-first} present the change in activities over time. We can see from the figures that both commit and discussion activities remain stable before and after the intervention.
However, during the unstable period, developers tend to be more active than usual.
In response to this phenomenon, we analyzed the persistent and transient effects of the interventions using the ITS method and Wilcoxon paired test method, respectively.

\subsubsection{ITS analysis}
Table \ref{result-its} shows the results of the ITS analysis.
The results show that the factor with the strongest correlation to OSS activity is the associated historical activity (\emph{i.e., number of commits before} for \textbf{Commit Model}, \emph{number of discussions before} for \textbf{Discussion Model}). For all four models, the associated historical activity explains more than 80\% of the total variance.
For the impact of other funding sources, we find that the variance explained by this factor does not exceed 1.1\% in all four models.
Therefore, it is somewhat clear that the existence of funding sources other than the \textsc{Sponsor} mechanism does not influence our exploration of the association of this mechanism with open source activity.

\begin{table*}[htbp] \centering
\setlength\tabcolsep{2.5pt}
\renewcommand\arraystretch{1.00}
\scriptsize
\caption{Results of ITS analysis} 
\label{result-its} 
\begin{tabular}{p{0.25\linewidth} p{0.08\linewidth}<{\centering} p{0.08\linewidth}<{\centering} p{0.08\linewidth}<{\centering} p{0.08\linewidth}<{\centering} p{0.08\linewidth}<{\centering} p{0.08\linewidth}<{\centering} p{0.08\linewidth}<{\centering} p{0.08\linewidth}<{\centering}}
\toprule
& \multicolumn{4}{c}{\textbf{Commit Model}} & \multicolumn{4}{c}{\textbf{Discussion Model}} \\ 
& \multicolumn{4}{c}{\textit{Dependent variable: scale(log(number of commits + 0.5))}} & \multicolumn{4}{c}{\textit{Dependent variable: scale(log(number of discussions + 0.5))}} \\ 
\cmidrule(r){2-5}  \cmidrule(r){6-9}
& \multicolumn{2}{c}{\emph{accountSetUpTime}} & \multicolumn{2}{c}{\emph{firstSponsorTime}} & \multicolumn{2}{c}{\emph{accountSetUpTime}} & \multicolumn{2}{c}{\emph{firstSponsorTime}} \\
\cmidrule(r){2-3} \cmidrule(r){4-5} \cmidrule(r){6-7} \cmidrule(r){8-9}
& Coeffs (Err.) & Chisq & Coeffs (Err.) & Chisq & Coeffs (Err.) & Chisq & Coeffs (Err.) & Chisq \\
\hline
(Intercept) & $-0.10^{***}(0.01)$ & & $-0.13^{***}(0.03)$ & & $\,\,\,0.01^{\,\,\,\,\,\,\,\,\,}(0.01)$ & & $-0.02^{\,\,\,\,\,\,\,\,\,}(0.04)$ & \\
 scale(log(\emph{number of commits before} + 0.5)) & $\,\,\,0.59^{***}(0.01)$ & $5190.72^{***}$ & $\,\,\,0.58^{***}(0.02)$ & $1185.38^{***}$ & $\,\,\,0.03^{***}(0.01)$ & $\,\,\,\,\,\,16.78^{***}$ & $-0.02^{\,\,\,\,\,\,\,\,\,}(0.02)$ & $\,\,\,\,\,\,\,\,\,1.14^{\,\,\,\,\,\,\,\,\,}$ \\
 scale(log(\emph{number of discussions before} + 0.5)) & $-0.02^{.\,\,\,\,\,\,}(0.01)$ & $\,\,\,\,\,\,\,\,\,3.45^{*\,\,\,\,\,\,}$ & $-0.03^{\,\,\,\,\,\,\,\,\,}(0.02)$ & $\,\,\,\,\,\,\,\,\,2.29^{\,\,\,\,\,\,\,\,\,}$ & $\,\,\,0.57^{***}(0.01)$ & $4570.67^{***}$ & $\,\,\,0.60^{***}(0.02)$ & $1333.36^{***}$ \\
 scale(log(\emph{number of stars before} + 0.5)) & $-0.06^{***}(0.01)$ & $\,\,\,\,\,\,55.23^{***}$ & $-0.07^{***}(0.01)$ & $\,\,\,\,\,\,22.71^{***}$ & $-0.02^{**\,\,\,}(0.01)$ & $\,\,\,\,\,\,\,\,\,6.41^{*\,\,\,\,\,\,}$ & $-0.05^{***}(0.01)$ & $\,\,\,\,\,\,11.11^{***}$ \\
 \emph{has goal} (TRUE) & $\,\,\,0.06^{***}(0.01)$ & $\,\,\,\,\,\,17.43^{***}$ & $\,\,\,0.07^{*\,\,\,\,\,\,}(0.03)$ & $\,\,\,\,\,\,\,\,\,5.97^{*\,\,\,\,\,\,}$ & $\,\,\,0.01^{\,\,\,\,\,\,\,\,\,}(0.01)$ & $\,\,\,\,\,\,\,\,\,1.02^{\,\,\,\,\,\,\,\,\,}$ & $\,\,\,0.03^{\,\,\,\,\,\,\,\,\,}(0.03)$ & $\,\,\,\,\,\,\,\,\,1.21^{\,\,\,\,\,\,\,\,\,}$ \\
 \emph{has other way} (TRUE) & $\,\,\,0.16^{**\,\,\,}(0.05)$ & $\,\,\,\,\,\,\,\,\,8.22^{**\,\,\,}$ & $\,\,\,0.14^{\,\,\,\,\,\,\,\,\,}(0.09)$ & $\,\,\,\,\,\,\,\,\,2.36^{\,\,\,\,\,\,\,\,\,}$ & $\,\,\,0.28^{***}(0.05)$ & $\,\,\,\,\,\,26.17^{***}$ & $\,\,\,0.14^{\,\,\,\,\,\,\,\,\,}(0.09)$ & $\,\,\,\,\,\,\,\,\,2.33^{\,\,\,\,\,\,\,\,\,}$ \\
 \emph{in company} (TRUE) & $\,\,\,0.09^{***}(0.01)$ & $\,\,\,\,\,\,38.56^{***}$ & $\,\,\,0.11^{***}(0.03)$ & $\,\,\,\,\,\,15.60^{***}$ & $\,\,\,0.01^{\,\,\,\,\,\,\,\,\,}(0.01)$ & $\,\,\,\,\,\,\,\,\,0.31^{\,\,\,\,\,\,\,\,\,}$ & $\,\,\,0.02^{\,\,\,\,\,\,\,\,\,}(0.03)$ & $\,\,\,\,\,\,\,\,\,0.60^{\,\,\,\,\,\,\,\,\,}$ \\
 \emph{is hireable} (TRUE) & $\,\,\,0.00^{\,\,\,\,\,\,\,\,\,}(0.01)$ & $\,\,\,\,\,\,\,\,\,0.02^{\,\,\,\,\,\,\,\,\,}$ & $\,\,\,0.01^{\,\,\,\,\,\,\,\,\,}(0.03)$ & $\,\,\,\,\,\,\,\,\,0.22^{\,\,\,\,\,\,\,\,\,}$ & $-0.08^{***}(0.01)$ & $\,\,\,\,\,\,30.34^{***}$ & $-0.06^{*\,\,\,\,\,\,}(0.03)$ & $\,\,\,\,\,\,\,\,\,4.41^{*\,\,\,\,\,\,}$ \\
 \emph{time} & $\,\,\,0.02^{***}(0.00)$ & $\,\,\,\,\,\,96.11^{***}$ & $\,\,\,0.03^{***}(0.00)$ & $\,\,\,\,\,\,61.22^{***}$ & $\,\,\,0.01^{***}(0.00)$ & $\,\,\,\,\,\,21.42^{***}$ & $\,\,\,0.02^{***}(0.00)$ & $\,\,\,\,\,\,21.80^{***}$ \\
 \emph{intervention} (TRUE) & $-0.02^{*\,\,\,}(0.01)$ & $\,\,\,\,\,\,\,\,\,5.66^{*\,\,\,\,\,\,}$ & $-0.09^{***}(0.02)$ & $\,\,\,\,\,\,25.54^{***}$ & $\,\,\,0.01^{\,\,\,\,\,\,\,\,\,}(0.01)$ & $\,\,\,\,\,\,\,\,\,0.30^{\,\,\,\,\,\,\,\,\,}$ & $-0.05^{**\,\,\,}(0.02)$ & $\,\,\,\,\,\,\,\,\,6.71^{**\,\,\,}$ \\
 \emph{time after intervention} & $-0.04^{***}(0.00)$ & $\,\,\,245.92^{***}$ & $-0.05^{***}(0.00)$ & $\,\,\,\,\,\,97.38^{***}$ & $-0.03^{***}(0.00)$ & $\,\,\,111.52^{***}$ & $-0.04^{***}(0.00)$ & $\,\,\,\,\,\,63.73^{***}$ \\
\hline \\
[-3ex]
Number of Observations & \multicolumn{2}{c}{75,516} & \multicolumn{2}{c}{20,148} & \multicolumn{2}{c}{75,516} & \multicolumn{2}{c}{20,148} \\
$R^{2}_m$ & \multicolumn{2}{c}{0.34} & \multicolumn{2}{c}{0.32} & \multicolumn{2}{c}{0.37} & \multicolumn{2}{c}{0.35} \\
$R^{2}_c$ & \multicolumn{2}{c}{0.64} & \multicolumn{2}{c}{0.64} & \multicolumn{2}{c}{0.66} & \multicolumn{2}{c}{0.65} \\
\bottomrule
\multicolumn{9}{l}{$^{***}p<0.001, ^{**}p<0.01, ^{*}p<0.05, ^{.}<0.1$}
\end{tabular}
\end{table*}

For the \emph{number of commits},
we find that for both \emph{accountSetUpTime} and \emph{firstSponsorTime}, there is a slight growth trend before the intervention. After the intervention, both show a negative growth trend ($\beta(time)+\beta(time\ after\ intervention)<0$).
Additionally, we find that the intervention itself is negatively correlated with the \emph{number of commits} ($\beta(intervention)<0$).

For the \emph{number of discussions},
we find results similar to those for the commit activity. The intervention of the \textsc{Sponsor} mechanism changes the original slowly increasing dynamics and reduces the discussion activity.
Specifically, the intervention has no effect at \emph{accountSetUpTime} but a slightly negative effect at \emph{firstSponsorTime}.

In regard to the above results, it is surprising that the setup of the \textsc{Sponsor} mechanism or the first sponsorship does not contribute to the \emph{maintainer}'s commit activity or discussion activity growth. In contrast, there is a slight inhibitory effect.
To illuminate this situation, we followed up with a questionnaire to explore the \emph{maintainers'} subjective satisfaction with the \textsc{Sponsor} mechanism and its motivating effect (see Section \ref{questionnaire-survey-RQ2}).

\subsubsection{Wilcoxon paired test analysis}
Table \ref{result-wilcoxon} shows the results of the Wilcoxon paired test and Cliff's delta.
\begin{table*}[htbp] \centering
  \setlength\tabcolsep{2.5pt}
  \renewcommand\arraystretch{1.00}
  \scriptsize
  \caption{Results of Wilcoxon paired test}
  \label{result-wilcoxon} 
\begin{tabular}{p{0.10\linewidth} p{0.06\linewidth}<{\centering} p{0.14\linewidth}<{\centering} p{0.14\linewidth}<{\centering} p{0.09\linewidth}<{\centering} p{0.14\linewidth}<{\centering} p{0.14\linewidth}<{\centering} p{0.09\linewidth}<{\centering}}
\toprule
 & & \multicolumn{3}{c}{\textbf{Commit}} & \multicolumn{3}{c}{\textbf{Discussion}} \\
 \cmidrule(r){3-5}  \cmidrule(r){6-8}
 & Num obs & Wilcoxon(\emph{greater}) p value & Wilcoxon(\emph{less}) p value & Cliff's delta & Wilcoxon(\emph{greater}) p value & Wilcoxon(\emph{less}) p value & Cliff's delta  \\
\hline
  \emph{accountSetUpTime} & 7,969 & $8e-05^{***}$ & $0.9999^{\,\,\,\,\,\,\,\,\,}$ & $\,\,\,0.022$ & $0.6801^{\,\,\,\,\,\,\,\,\,}$ & $0.3199^{\,\,\,\,\,\,\,\,\,}$ & $-0.005$ \\
  \emph{firstSponsorTime} & 2,796 & $\,\,\,0.0015^{**\,\,\,}$ & $0.9985^{\,\,\,\,\,\,\,\,\,}$ & $\,\,\,0.025$ & $0.7433^{\,\,\,\,\,\,\,\,\,}$ & $0.2567^{\,\,\,\,\,\,\,\,\,}$ & $-0.009$ \\
  all sponsored time & 21,153 & $3e-12^{***}$ & $1.0000^{\,\,\,\,\,\,\,\,\,}$ & $\,\,\,0.021$ & $0.9809^{\,\,\,\,\,\,\,\,\,}$ & $0.0191^{\,\,\,\,\,\,\,\,\,}$ & $-0.003$ \\
\bottomrule
\multicolumn{5}{l}{$^{***}p<0.001, ^{**}p<0.01, ^{*}p<0.05, ^{.}<0.1$}
\end{tabular}
\end{table*}


For the \emph{number of commits}, when the \emph{maintainer} sets up the \textsc{Sponsor} account, is sponsored for the first time, or receives a new sponsorship, the \emph{number of commits} after the intervention is significantly higher.
For the \emph{number of discussions}, we find no significant changes around the three kinds of interventions.

This result indicates that sponsor behavior leads to a short-term increase in commit activity.
For discussion, however, the sponsorship does not lead to short-term changes.
In contrast to the ITS analysis, the Wilcoxon paired test analyzes changes in activity during the unstable period, further demonstrating that the \textsc{Sponsor} mechanism can give a short-term boost to development activity.

\subsubsection{Questionnaire survey}
\label{questionnaire-survey-RQ2}
\begin{figure}
  \centering
  \includegraphics[width=\linewidth]{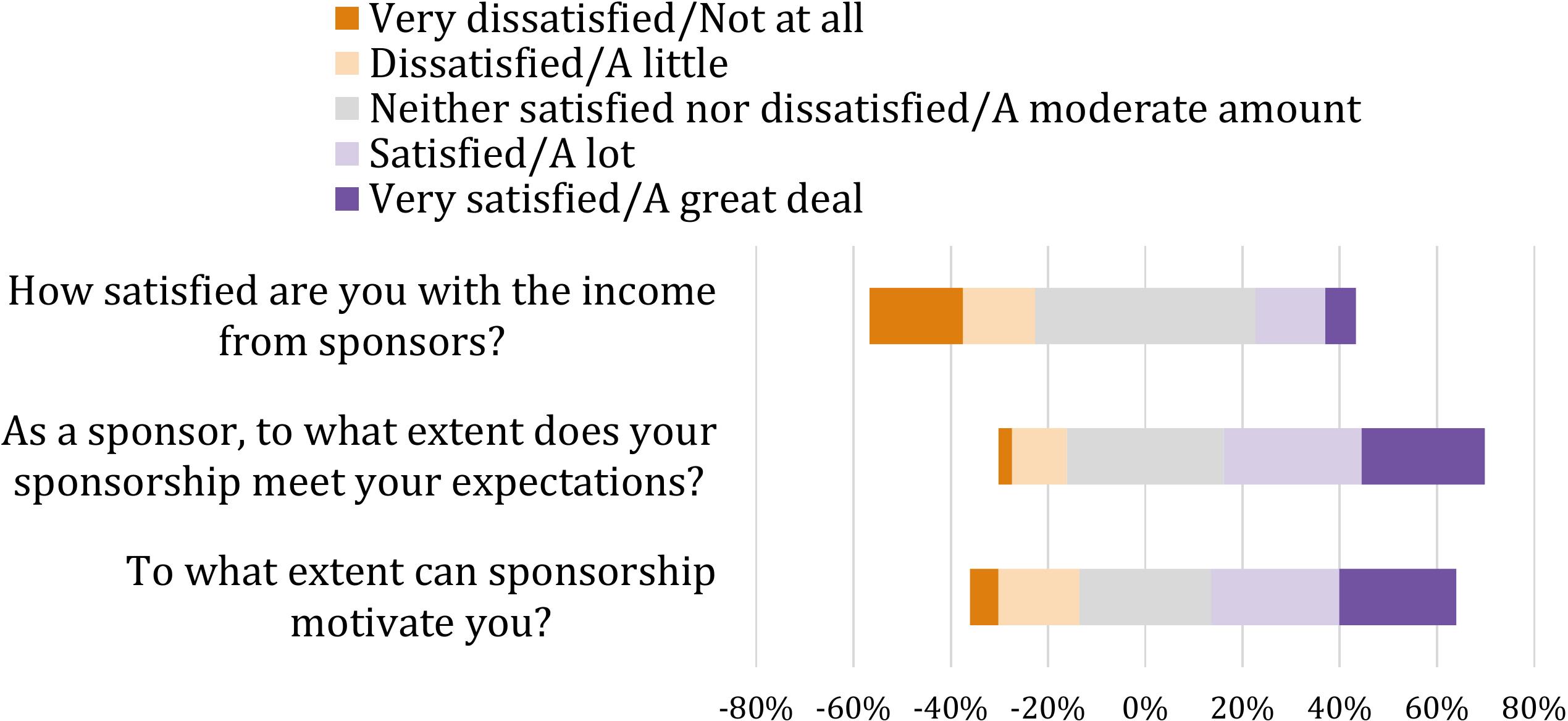}
\caption{Results of 5-point Likert scale questions}
  \label{fig:likert-scale-results}
  \Description{This figure presents the result of how maintainers answer Q4, Q5, and how sponsors answer Q4. It shows that more maintainers are not satisfied with the income, while more sponsors think the \textsc{Sponsor} mechanism meets their expectations and more maintainers think the \textsc{Sponsor} mechanism can motivate them.}
\end{figure}

To further explore the effectiveness of the \textsc{Sponsor} mechanism, we conducted independent research with \emph{maintainers} and \emph{sponsors} to uncover their subjective judgments about the efficacy of the mechanism.
In response to this goal, we asked \emph{\emph{maintainers}} (\textbf{Q4 ``How satisfied are you with the income from sponsors?''}) and \emph{\emph{sponsors}} (\textbf{Q4 ``As a sponsor, to what extent does your sponsorship meet your expectations?''}).
Meanwhile, we asked the maintainers directly about their internal perceptions of the effectiveness of sponsorship incentives (\textbf{Q5 ``To what extent can sponsorship motivate you?''}). The results are shown in Figure \ref{fig:likert-scale-results}.

For \emph{sponsors}, we find that 53.7\% think that sponsorship meets their expectations fully
or a great deal and only 14.1\% report that their expectations are hardly met or not met at all.
For \emph{maintainers}, we find that 50.4\% consider that sponsorship motivates them fully or a great deal but 22.5\% think that it does not bring any motivating effect.
However, in terms of the amount of sponsorship, we find that only 20.7\% of \emph{maintainers} are either satisfied or very satisfied with their income from sponsorship and 30.1\% are dissatisfied or very dissatisfied with the amount.

We think that the main reason for this difference is that sponsors' main motivation to participate is to display their gratitude, inspire others, etc., by giving funds. Therefore, most \emph{sponsors} are satisfied with their own behavior.
For \emph{maintainers}, although more than half think that sponsorship can be stimulating, we find that only approximately 20\% are satisfied with the amount of sponsorship received. This shows that open source sponsorship has a positive effect on some developers, but in fact, the amount of monetary rewards that can be received through sponsorship is relatively small and unlikely to meet the expectations of \emph{maintainers}.

\begin{tcolorbox}[left=0mm, right=0mm, top=0mm, bottom=0mm]
In terms of short-term effects, the \textsc{Sponsor} mechanism makes a slightly positive contribution to the development activity but has no significant impact on discussion activity.
However, this impact is not sustained.
One possible reason is that
the actual amount of support does not meet \emph{maintainers'} expectations, which makes it difficult for \emph{maintainers} to rely on sponsorship income to keep investing in open source contributions.
\end{tcolorbox}

\subsection{RQ3: \textit{Who} is likely to receive more sponsorships?}

For this research question, we tried to identify the important factors influencing the amount of sponsorship and provide further advice on \emph{maintainers}.
We again analyzed and verified the results through a combination of quantitative and qualitative analysis.
For the qualitative analysis, we analyzed both \emph{maintainers} and \emph{sponsors} and explored the consistency of their perceptions of sponsorship.

\subsubsection{Hurdle regression}
\begin{table*}[htbp] \centering
    \setlength\tabcolsep{2.5pt}
    \renewcommand\arraystretch{1.00}
    \footnotesize
    \caption{Result for factors influencing sponsorship} 
    \label{result-sponsorship} 
  \begin{tabular}{p{0.28\linewidth} p{0.17\linewidth}<{\centering} p{0.17\linewidth}<{\centering} p{0.17\linewidth}<{\centering} p{0.17\linewidth}<{\centering}}
  \toprule
   & \multicolumn{2}{c}{\textit{Dependent variable: receive sponsorship}} & \multicolumn{2}{c}{\textit{Dependent variable: the amount of sponsorship}} \\ 
   \cmidrule(r){2-3}  \cmidrule(r){4-5}
   & Coeffs (Err.) & Chisq & Coeffs (Err.) & Chisq \\
  \hline
  (Intercept) & $-0.53^{***}(0.09)$ & & $\,\,\,1.80^{***}(0.07)$ & \\
  scale(log(\emph{user age} + 0.5)) & $-0.10^{**\,\,\,}(0.03)$ & $\,\,\,\,\,\,8.62^{**\,\,\,}$ & $-0.02^{\,\,\,\,\,\,\,\,\,}(0.02)$ & $\,\,\,\,\,\,0.41^{\,\,\,\,\,\,\,\,\,}$ \\
  \emph{in company} (TRUE) & $-0.26^{***}(0.06)$ & $\,\,\,18.08^{***}$ & $-0.12^{**\,\,\,}(0.04)$ & $\,\,\,\,\,\,7.48^{**\,\,\,}$ \\
  \emph{has email} (TRUE) & $-0.03^{\,\,\,\,\,\,\,\,\,}(0.06)$ & $\,\,\,\,\,\,0.31^{\,\,\,\,\,\,\,\,\,}$ & $\,\,\,0.12^{**\,\,\,}(0.04)$ & $\,\,\,\,\,\,7.53^{**\,\,\,}$ \\
  \emph{has location} (TRUE) & $-0.11^{\,\,\,\,\,\,\,\,\,}(0.09)$ & $\,\,\,\,\,\,1.41^{\,\,\,\,\,\,\,\,\,}$ & $-0.19^{**\,\,\,}(0.06)$ & $\,\,\,\,\,\,8.36^{**\,\,\,}$ \\
  \emph{is hireable} (TRUE) & $-0.19^{**\,\,\,}(0.06)$ & $\,\,\,\,\,\,9.70^{**\,\,\,}$ & $-0.07^{\,\,\,\,\,\,\,\,\,}(0.04)$ & $\,\,\,\,\,\,2.28^{\,\,\,\,\,\,\,\,\,}$ \\
  scale(log(\emph{followers} + 0.5)) & $\,\,\,0.96^{***}(0.04)$ & $545.36^{***}$ & $\,\,\,0.68^{***}(0.03)$ & $722.89^{***}$ \\
  scale(log(\emph{followings} + 0.5)) & $-0.19^{***}(0.03)$ & $\,\,\,37.39^{***}$ & $-0.24^{***}(0.02)$ & $118.53^{***}$ \\
  scale(log(\emph{min tier} + 0.5)) & $-0.42^{***}(0.04)$ & $146.89^{***}$ & $-0.04^{.\,\,\,\,\,\,\,}(0.02)$ & $\,\,\,\,\,\,3.39^{.\,\,\,\,\,\,\,}$ \\
  scale(log(\emph{max tier} + 0.5)) & $\,\,\,0.23^{***}(0.03)$ & $\,\,\,59.82^{***}$ & $\,\,\,0.14^{***}(0.02)$ & $\,\,\,38.46^{***}$ \\
  \emph{has goal} (TRUE) & $\,\,\,0.18^{**\,\,\,}(0.06)$ & $\,\,\,\,\,\,8.32^{**\,\,\,}$ & $\,\,\,0.01^{\,\,\,\,\,\,\,\,\,}(0.04)$ & $\,\,\,\,\,\,0.10^{\,\,\,\,\,\,\,\,\,}$ \\
  \emph{has other way} (TRUE) & $\,\,\,0.28^{\,\,\,\,\,\,\,\,\,}(0.22)$ & $\,\,\,\,\,\,1.54^{\,\,\,\,\,\,\,\,\,}$ & $\,\,\,0.44^{***}(0.13)$ & $\,\,\,13.51^{***}$ \\
  scale(log(\emph{user age after sponsor account} + 0.5)) & $\,\,\,0.02^{\,\,\,\,\,\,\,\,\,}(0.03)$ & $\,\,\,\,\,\,0.40^{\,\,\,\,\,\,\,\,\,}$ & $-0.02^{\,\,\,\,\,\,\,\,\,}(0.02)$ & $\,\,\,\,\,\,0.81^{\,\,\,\,\,\,\,\,\,}$ \\
  scale(log(\emph{number of commits} + 0.5)) & $\,\,\,0.08^{.\,\,\,\,\,\,\,}(0.04)$ & $\,\,\,\,\,\,3.42^{.\,\,\,\,\,\,\,}$ & $\,\,\,0.10^{***}(0.03)$ & $\,\,\,10.68^{**\,\,\,}$ \\
  scale(log(\emph{number of discussions} + 0.5)) & $\,\,\,0.73^{***}(0.05)$ & $270.29^{***}$ & $\,\,\,0.31^{***}(0.03)$ & $106.73^{***}$ \\
  scale(log(\emph{sum star number} + 0.5)) & $-0.10^{**\,\,\,}(0.04)$ & $\,\,\,\,\,\,7.48^{**\,\,\,}$ & $-0.07^{**\,\,\,}(0.03)$ & $\,\,\,\,\,\,6.87^{**\,\,\,}$ \\
  scale(log(\emph{sum top repository star number} + 0.5)) & $-0.13^{**\,\,\,}(0.04)$ & $\,\,\,\,\,\,9.55^{**\,\,\,}$ & $-0.15^{***}(0.03)$ & $\,\,\,31.11^{***}$ \\
  scale(log(\emph{introduction richness} + 0.5)) & $\,\,\,0.23^{***}(0.03)$ & $\,\,\,60.84^{***}$ & $\,\,\,0.11^{***}(0.02)$ & $\,\,\,25.74^{***}$ \\
  scale(log(\emph{number of dependents} + 0.5)) & $-0.02^{\,\,\,\,\,\,\,\,\,}(0.03)$ & $\,\,\,\,\,\,0.62^{\,\,\,\,\,\,\,\,\,}$ & $-0.03^{.\,\,\,\,\,\,\,}(0.02)$ & $\,\,\,\,\,\,2.69^{\,\,\,\,\,\,\,\,\,}$ \\
   \hline \\
   [-3ex]
  Number of Observations & \multicolumn{2}{c}{7,465} & \multicolumn{2}{c}{2,750} \\
  delta $R^{2}$ & \multicolumn{2}{c}{0.34} & \multicolumn{2}{c}{0.39} \\
  \bottomrule
  \multicolumn{5}{l}{$^{***}p<0.001, ^{**}p<0.01, ^{*}p<0.05, ^{.}<0.1$}
  \end{tabular}
  \end{table*}

From an overall perspective (see Table \ref{result-sponsorship}), the hurdle regression models fit well, with $R^{2}=34\%$ and $R^{2}=39\%$, respectively.
Even though 7,465 \emph{maintainers} have more than 3 months of activity after setting up their \textsc{Sponsor} profile, only 2,750 (36.8\%) of them receive at least one sponsorship.
Moreover, only 6\% receive sponsorships more than 10 times, and only 25 \emph{maintainers} receive more than 100 sponsorships.
Therefore, although many people want to obtain sponsorship, only a small number of people succeed.

When we consider whether the \emph{maintainer} receives any sponsorships (columns 2 and 3 of Table~\ref{result-sponsorship}),
the \emph{followers} factor, representing social status, has the most substantial positive effect, explaining 45.8\% of the total variance.
However, the factor \emph{followings} is negatively correlated with the likelihood of receiving sponsorship (effect size: 3.1\%). It is likely that compared to \emph{followings}, \emph{followers} better represents the centrality of \emph{maintainers} in the community, while \emph{maintainers} with large \emph{followings} tend to learn more from others in the community.
Discussion activity is positively correlated with the likelihood of sponsorship (\emph{number of discussions}, effect size: 22.7\%), while relatively speaking, commit activity explains only 0.3\% of the variance. A possible explanation is that sponsored developers tend to focus more on issues or pull requests submitted by sponsors to give back or attract the attention of others. Commit activity is common among GitHub developers, where many developers may just focus on their own issues.
For sponsor tiers, the \emph{min tier} is negatively correlated with the likelihood of sponsorship acquisition (effect size: 12.3\%). However, \emph{max tier} is positively correlated and explains 5\% of the variance.
Both of the tiers have sizable effects but opposite directions of influence.
It is likely that many sponsors tend to donate only a little money and that setting a high \emph{min tier} may cause them to abstain from sponsorship.
However, if \emph{maintainers} want to obtain sponsorships, they cannot undervalue themselves. Trying to increase the \emph{max tier} can increase the possibility of being sponsored.
Another thing for \emph{maintainers} to note is the importance of the introduction text when setting up their \textsc{Sponsor} account. If \emph{maintainers} introduce themselves at greater length, they are more likely to become sponsored (effect size: 5.1\%).
Other factors have negligible effects, with explained variances of less than 5\%.

When we consider the amount of sponsorship received by \emph{maintainers} (columns 4 and 5 of Table~\ref{result-sponsorship}), the social status of \emph{maintainers} is also positively correlated with the response (\emph{followers}, effect size: 65.3\%). At the same time, \emph{followings} oppositely correlates with the response (effect size: 10.7\%).
The factor \emph{number of discussions} explains 9.6\% of the total variance.
The \emph{min tier} variable becomes nonsignificant, unlike in the \emph{receive sponsorship} model. A possible explanation for this result is that the setting of the \emph{min tier} is not a long-term solution for securing more sponsorship. Developers need to be more focused on their status and daily activities in the community.
Other factors have negligible effects.

\subsubsection{Questionnaire}
We asked questions related to \emph{maintainers} (\textbf{Q6 ``In which way do you think you can obtain more sponsorships?''}) and \emph{sponsors} (\textbf{Q5 ``What kind of developer do you prefer to sponsor?''}) separately.
Table \ref{result-ways} presents the results.
\begin{table*}[htbp] \centering
    \setlength\tabcolsep{2.5pt}
    \renewcommand\arraystretch{1.00}
    \footnotesize
    \caption{Ways of obtaining more sponsorship}
    \label{result-ways} 
  \begin{tabular}{p{7.0cm}<{\raggedright}p{0.5cm}<{\raggedleft}p{1.1cm}<{\raggedright}p{6.4cm}<{\raggedright}p{0.5cm}<{\raggedleft}p{1.1cm}<{\raggedright}}
  \toprule
   \textbf{Way\_maintainers} & \multicolumn{2}{l}{\textbf{Votes (\%)}} & \textbf{Who\_sponsors} & \multicolumn{2}{l}{\textbf{Votes (\%)}} \\
   \cmidrule(r){1-3}  \cmidrule(r){4-6}
   $\mathcal{WM}1$ Producing useful projects & 62.6  & \includegraphics[width=6.26mm, height=2mm]{pics/percentage.png} & $\mathcal{WS}1$ Developers whose projects I benefit from & 85.1 & \includegraphics[width=8.51mm, height=2mm]{pics/percentage.png} \\
   $\mathcal{WM}2$ Staying active and contributing more in the community & 54.5  & \includegraphics[width=5.45mm, height=2mm]{pics/percentage.png} & $\mathcal{WS}2$ Developers whose projects I’m interested in & 60.3 & \includegraphics[width=6.03mm, height=2mm]{pics/percentage.png} \\
   $\mathcal{WM}3$ Advertising myself or my work to the community & 54.1  & \includegraphics[width=5.41mm, height=2mm]{pics/percentage.png} & $\mathcal{WS}3$ Developers who make important contributions & 50.9 & \includegraphics[width=5.09mm, height=2mm]{pics/percentage.png} \\
   $\mathcal{WM}4$ Producing valuable code & 38.5  & \includegraphics[width=3.85mm, height=2mm]{pics/percentage.png} & $\mathcal{WS}4$ Developers who are active in community & 42.0 & \includegraphics[width=4.2mm, height=2mm]{pics/percentage.png} \\
   $\mathcal{WM}5$ Getting involved in popular projects & 29.1  & \includegraphics[width=2.91mm, height=2mm]{pics/percentage.png} & $\mathcal{WS}5$ Independent developers & 31.1 & \includegraphics[width=3.11mm, height=2mm]{pics/percentage.png} \\
   $\mathcal{WM}6$ Getting involved in projects adopted by companies & 25.5  & \includegraphics[width=2.55mm, height=2mm]{pics/percentage.png} & $\mathcal{WS}6$ Developers who haven’t received much sponsorship & 24.1 & \includegraphics[width=2.41mm, height=2mm]{pics/percentage.png} \\
   $\mathcal{WM}7$ Getting involved in long-term projects & 21.6  & \includegraphics[width=2.16mm, height=2mm]{pics/percentage.png} & $\mathcal{WS}7$ Developers who are in hardship & 18.7 & \includegraphics[width=1.87mm, height=2mm]{pics/percentage.png} \\
   $\mathcal{WM}8$ Getting involved in less maintained yet important projects & 19.1  & \includegraphics[width=1.91mm, height=2mm]{pics/percentage.png} & $\mathcal{WS}8$ Developers who I know & 15.4 & \includegraphics[width=1.54mm, height=2mm]{pics/percentage.png} \\
   $\mathcal{WM}9$ Getting involved in projects led by companies & 8.8   & \includegraphics[width=0.88mm, height=2mm]{pics/percentage.png} & $\mathcal{WS}9$ Other & 1.0 & \includegraphics[width=0.1mm, height=2mm]{pics/percentage.png} \\
   $\mathcal{WM}10$ Providing localized content & 7.4   & \includegraphics[width=0.74mm, height=2mm]{pics/percentage.png} &       &       &       \\
   $\mathcal{WM}11$ Other & 3.6   & \includegraphics[width=0.36mm, height=2mm]{pics/percentage.png} &       &       &       \\

  \bottomrule
  \end{tabular}
  \end{table*}

\paragraph{For \emph{maintainers}}
The results reveal that from the \emph{maintainers'} perspective, producing useful projects and tools ($\mathcal{WM}1$, $\mathcal{WM}4$) is seen as more likely to draw sponsorships than just participating in projects 
($\mathcal{WM}5$, $\mathcal{WM}6$, $\mathcal{WM}7$, $\mathcal{WM}8$, $\mathcal{WM}9$).
One possible reason for this is that the \textsc{Sponsor} mechanism is to credit funds to individual accounts, and the sponsorship button on the project homepage also needs to be configured by the owner. Some \emph{sponsors} who want to donate to a project through the \textsc{Sponsor} mechanism (\emph{e.g., those reporting that ``I prefer to sponsor projects, not a specific developer'' [SC167]}) may end up sponsoring only the project's owner.

Some 54.5\% of \emph{maintainers} think that by working hard, they can obtain more sponsorships (\textbf{$\mathcal{WM}2$}). However, some \emph{maintainers} said sponsorship is simply a matter of popularity (\emph{e.g., ``Purely popularity basically... OSS Creators from YouTube earn a ton of money'' [MC292]; ``I think it is mostly a function of being a celebrity so it operates on the same rules'' [MC262]}). This is probably why 54.1\% of the \emph{maintainers} chose \textbf{$\mathcal{WM}3$}.

More than 1 option was chosen by 85.6\% of the sponsored participants. Moreover, 20.5\% chose at least 5 options, which shows that in fact, the options that we offered are feasible for promoting sponsorships among \emph{maintainers}. Some relevant participants indicated that \emph{``Donations just don't work'' [MC284] or ``It doesn't matter; people take when it's free'' [MC281]}.
These responses suggest that the reasons that prevent most people from obtaining more sponsorships that would meet their expectations are not limited to individual participation characteristics and platform mechanism design; rather, the act of sponsorship itself may not be suitable for the open source sphere. Indeed, 10 participants who selected \textbf{$\mathcal{WM}11$} indicated that there was no way to obtain more sponsorship.

\paragraph{For \emph{sponsors}}
The vast majority (85.1\%) chose \textbf{$\mathcal{WS}1$}, which suggests that most \emph{sponsors} support developers involved in the open source projects that sponsors use.
This corresponds to the top-ranked way of obtaining sponsorship (\textbf{$\mathcal{WM}1$}) selected by the \emph{maintainers}, suggesting that the best way to obtain more sponsorship, in the opinion of both \emph{maintainers} and \emph{sponsors}, is to create projects that more people use.
Similarly, more than half of the participants wanted to sponsor projects of personal interest (\textbf{$\mathcal{WS}2$}) and developers who had made significant contributions (\textbf{$\mathcal{WS}3$}).
We find that 31.1\% of the \emph{sponsors} chose to sponsor independent developers (\textbf{$\mathcal{WS}5$}). However, some \emph{sponsors} said that just being an independent developer is not enough and that the development and maintenance of good open source projects or tools are needed (\emph{e.g., ``Independent developers with nice tools'' [SC30]}).

Most \emph{sponsors} do not consider the act of sponsorship as a form of charity---few people reported doing so simply because the person being rewarded was in hardship (\textbf{$\mathcal{WS}7$}) or had not received many rewards (\textbf{$\mathcal{WS}6$}).
Likewise, \emph{sponsors} do not want to reward another developer simply because they know one another (only 15.4\% chose \textbf{$\mathcal{WS}8$}; \emph{e.g., ``It is usually a library I am using in my own project and I know the developer in person'' [SC168])}.

\begin{tcolorbox}[left=0mm, right=0mm, top=0mm, bottom=0mm]
Most maintainers and sponsors think that sponsorship builds on relationships forged through using OSS.
Active and meaningful participation in open source contributions can also help maintainers gain more attention.
However, the quantitative analysis reveals that the social popularity of the \emph{maintainer} in the community is the decisive factor in obtaining more sponsorships.
\end{tcolorbox}

\subsection{RQ4: \textit{What} are the shortcomings of the \textsc{Sponsor} mechanism?}
\label{Result-RQ4}

For this research question, we investigated the mechanism shortcomings found by participants while using the \textsc{Sponsor} mechanism.
We asked the question ``\textbf{What are the shortcomings of the \textsc{Sponsor} mechanism?}'' of both maintainers (\textbf{Q7}) and sponsors (\textbf{Q6}) separately.
Table \ref{result-shortcomings} presents the results.

\begin{table*}[htbp] \centering
    \setlength\tabcolsep{1pt}
    \renewcommand\arraystretch{1.00}
    \footnotesize
    \caption{Shortcomings of the \textsc{Sponsor} mechanism}
    \label{result-shortcomings} 
  \begin{tabular}{p{7.2cm}<{\raggedright}p{0.5cm}<{\raggedleft}p{0.9cm}<{\raggedright}p{7cm}<{\raggedright}p{0.5cm}<{\raggedleft}p{0.9cm}<{\raggedright}}
  \toprule
   \textbf{Shortcoming\_maintainers} & \multicolumn{2}{l}{\textbf{Votes (\%)}} & \textbf{Shortcoming\_sponsors} & \multicolumn{2}{l}{\textbf{Votes (\%)}} \\
   \cmidrule(r){1-3}  \cmidrule(r){4-6}
   
   $\mathcal{SM}1$ It’s hard for others to discover me for sponsorship & 51.3  & \includegraphics[width=5.13mm, height=2mm]{pics/percentage.png} & $\mathcal{SS}1$ I cannot assess how urgently a developer needs to be sponsored & 40.1 & \includegraphics[width=4.01mm, height=2mm]{pics/percentage.png} \\
   $\mathcal{SM}2$ I can’t interact with my sponsors on GitHub (e.g., for expressing appreciation) & 29.4  & \includegraphics[width=2.94mm, height=2mm]{pics/percentage.png} & $\mathcal{SS}2$ None. It’s perfect & 33.1 & \includegraphics[width=3.31mm, height=2mm]{pics/percentage.png} \\
   $\mathcal{SM}3$ Lack of a wide range of payment options (e.g., one-time/yearly/quarterly payment) & 25.1  & \includegraphics[width=2.51mm, height=2mm]{pics/percentage.png} & $\mathcal{SS}3$ It’s hard for me to find the developer I should sponsor & 19.6 & \includegraphics[width=1.96mm, height=2mm]{pics/percentage.png} \\
   $\mathcal{SM}4$ GitHub does not distinctly mark my sponsors (e.g., I cannot easily tell whether an issue submitter is my sponsor) & 20.7  & \includegraphics[width=2.07mm, height=2mm]{pics/percentage.png} & $\mathcal{SS}4$ It is not supported in many regions & 13.2 & \includegraphics[width=1.32mm, height=2mm]{pics/percentage.png} \\
   $\mathcal{SM}5$ I have to pay taxes & 19.3  & \includegraphics[width=1.93mm, height=2mm]{pics/percentage.png} & $\mathcal{SS}5$ I can’t interact with the developer I sponsored on GitHub & 11.8 & \includegraphics[width=1.18mm, height=2mm]{pics/percentage.png} \\
   $\mathcal{SM}6$ None. It's perfect to me & 13.1  & \includegraphics[width=1.31mm, height=2mm]{pics/percentage.png} & $\mathcal{SS}6$ I’m not distinctly marked in the projects whose maintainers have been sponsored by me (e.g., when I submit an issue) & 10.5 & \includegraphics[width=1.05mm, height=2mm]{pics/percentage.png} \\
   $\mathcal{SM}7$ It is not supported in many regions & 11.0  & \includegraphics[width=1.1mm, height=2mm]{pics/percentage.png} & $\mathcal{SS}7$ Other & 8.1 & \includegraphics[width=0.81mm, height=2mm]{pics/percentage.png} \\
   $\mathcal{SM}8$ I can’t declare how I dealt with the received money & 10.1  & \includegraphics[width=1.01mm, height=2mm]{pics/percentage.png} &       &       &  \\
   $\mathcal{SM}9$ Other & 9.4   & \includegraphics[width=0.94mm, height=2mm]{pics/percentage.png} &       &       &  \\

  \bottomrule
  \multicolumn{4}{l}{During the research process, GitHub fixed some shortcomings, \emph{e.g.,} the one-time payment method.}
  \end{tabular}
  \end{table*}

Among maintainers, 13.1\% thought that the \textsc{Sponsor} mechanism was perfect (\textbf{$\mathcal{SM}6$}) and could meet their personal needs well, while among sponsors, 33.1\% thought that the mechanism was perfect (\textbf{$\mathcal{SS}2$}). This indicates that the satisfaction of different types of mechanism participants, especially maintainers, varies greatly. The current \textsc{Sponsor} mechanism does not meet maintainers' needs well.
The shortcomings include the following main aspects (some of these were resolved by GitHub during the research process).

\paragraph{Discoverability of maintainers}
The results reveal that 51.3\% of maintainers found it difficult to be discovered by sponsors (\textbf{$\mathcal{SM}1$}); however, based on feedback from sponsors, only 19.6\% found it difficult to determine whom they should sponsor (\textbf{$\mathcal{SS}3$}). A larger share (40.1\%) found it difficult to assess who urgently needed sponsorship (\textbf{$\mathcal{SS}1$}).

\paragraph{Interactivity of participants}
From the results, we find that among maintainers, 29.4\% thought that the current \textsc{Sponsor} mechanism cannot support good direct communication with sponsors (\textbf{$\mathcal{SM}2$)}), while among sponsors, 11.8\% wanted communication support (\textbf{$\mathcal{SS}5$}). Some thought that they should not burden developers by interrupting their normal development process (\emph{``I don't want to burden the developers [by asking them] to communicate with sponsors. The sponsor should be string-free'' [SC195]}).

\paragraph{Payments}
Many people, including maintainers and sponsors, highlighted existing payment problems with the \textsc{Sponsor} mechanism, including limited payment options (25.1\% of maintainers -- \textbf{$\mathcal{SM}3$)}), limited sponsorship tiers, inconvenient tax payments (19.3\% of the maintainers -- \textbf{$\mathcal{SM}5$)}), and limited payment providers. Some of these shortcomings, \emph{e.g.,} the limited payment options, may have been resolved by GitHub during the research process.

\paragraph{User distinction}
A total of 20.7\% (\textbf{$\mathcal{SM}4$}) of maintainers and 10.5\% (\textbf{$\mathcal{SS}6$}) of sponsors mentioned the distinction between sponsors and others in project development activities.

\paragraph{Geographical restrictions} From \textbf{$\mathcal{SM}7$} and \textbf{$\mathcal{SS}4$}, we see that 11\% of maintainers and 13.2\% of sponsors thought that support for regions limits the popularity of participation. As of 27 July 2021, only 37 regions were supported, leaving many people unable to participate in the mechanism (\textbf{$\mathcal{RO}6$}) and \emph{sponsors} unable to sponsor as many people as they want (\emph{e.g., ``Not all organizations I want to support joined GitHub sponsors'' [SC192]}).

\paragraph{Lack of contribution indicators}
Five participants noted that there was a lack of valid OSS contribution indicators.
OSS contributions are not limited to commits and pull requests. If not involved in the current project, the \emph{sponsor} hardly knows who has played a significant role in the project development (\emph{e.g., ``It is not easy to measure my OSS contribution. Sometimes it is just filing issues; other times, it is documentation PRs'' [MC350]}).
Moreover, contributions of small patches to large projects are difficult for others to find and thus are unlikely to gain sponsorships (\emph{e.g., ``In my case, you will be hard-pressed to get anything for your work when you are making just a little addition to a massive piece of software'' [MC379]}).
Among sponsors, some want to sponsor a project, not individual maintainers (\emph{e.g., ``I prefer to sponsor projects, not a specific developer'' [SC167]}).

\paragraph{OSS donations}
The \textsc{Sponsor} mechanism itself is an act of donation. On GitHub, sponsorship is primarily for users or organizations that have created a GitHub account.
We find from the results that 16 participants thought that the donation mechanism itself was not suitable for the current open source sphere. Many reasons were cited for this evaluation: People take open source projects for granted, and no one wants to pay for them (\emph{e.g., ``People still do not like to pay for software'' [MC355]}). Companies that use open source initiatives to gain revenue do not want to give back to the open source project (\emph{e.g., ``Most companies don't fund any of their open source dependencies'' [MC354]}). Donations are passive income, and without a regular income, developers have little motivation to work full-time on open source projects (\emph{e.g., ``Donation makes far less revenue than charging for things'' [OC78]}).

To solve the problems mentioned above, we offer the following actionable suggestions after taking into account the participant feedback.

\paragraph{Discoverability of maintainers}
\begin{itemize}
\item Add ``Sponsor'' buttons for the relevant project or people on the release webpage (\emph{``Recognition of sponsors in release of the repository would be something I can think of'' [SC217]}).
\item Add support for integrated development environments (IDEs), allowing developers to discover package dependencies and quickly jump to sponsor pages while developing with IDEs (\emph{``Better discoverability and integration with other developer tooling'' [SC65]}).
\item Provide a more straightforward way to show personal OSS contributions (\emph{e.g., ``Promote efforts like a dashboard'' [MC126]}).
\end{itemize}

\paragraph{Interactivity among participants}
\begin{itemize}
\item Allow maintainers to configure themselves whether they wish to communicate directly with sponsors. The interaction can be set up in different groups for different sponsors, similar to Patreon's integration solution with Discord \cite{discord} (\emph{e.g., ``Lack of integration with the payment tiers like the Discord integration with Patreon'' [MC337]}).
\item Allow maintainers to configure their own thank-you emails that can be sent automatically when they receive a sponsorship (\emph{e.g., ``Some kind of thank-you setup where I can send notes, etc.'' [MC109]}).
\item Allow sponsors to upload statements to disclose expenses related to sponsorship proceeds (\emph{``Distribution of the money, especially in FOSS [free and open source software] projects'' [MC88]}).
\end{itemize}

\paragraph{Payments}
\begin{itemize}
\item Provide clear income and expense statements to the \emph{sponsor} and \emph{maintainer} automatically.
\item Integrate as many payment providers as possible on the basis of meeting tax requirements.
\end{itemize}

\paragraph{User distinctions}
\begin{itemize}
\item Let maintainers decide, through in a configurable form in their personal settings, whether they want to treat sponsors differently from nonsponsors.
\item In addition to an option to show distinctions, add configuration options such as what development activities to show and whether to distinguish between sponsors with different sponsorship amounts (\emph{e.g., ``Developers should be allowed to set permission levels based on sponsorship. E.g.,you can only comment or make requests if you're a sponsor (or if the developer directly opts you in, or if you've made contributions to the project, things like that). This would really positively change the culture of GitHub collaboration'' [SC212]}).
\end{itemize}

\paragraph{Geographical restrictions} Provide support for more regions.

\paragraph{Lack of contribution indicators} Set up a multidimensional indicator of contributions, and ensure rational allocation of project sponsorship funds.

\paragraph{OSS donations} Future research should synthesize feedback from all types of open source participants and reconsider how to improve the sponsorship mechanism or design a more appropriate form of open source financial support.

\begin{tcolorbox}[left=0mm, right=0mm, top=0mm, bottom=0mm]
  The shortcomings of the \textsc{Sponsor} mechanism relate to three main aspects.
  \emph{Usage deficiencies}: difficulty of participants in finding each other, lack of good interaction support, lack of promotion, lack of adequate payment and billing support, etc.
  \emph{Object orientation with supported functions}: despite support for organizations and projects, main targeting of individuals. For \emph{sponsors}, a need for better support for corporate sponsorship; for \emph{maintainers}, a need for better support for multicontributor projects.
  \emph{Personalization}: a need for configurability of the \textsc{Sponsor} mechanism to reflect variation in participant types and motivations.
\end{tcolorbox}

\section{Discussion}
\label{discussion}

Through this study of the integrated sponsorship mechanism on the world's most popular open source platform (GitHub), we found that participation in the mechanism has not shown the same rapid growth as participation in open source projects. Meanwhile, there is a long-tail effect regarding the number of sponsorships obtained by maintainers; \emph{i.e.,} most maintainers do not obtain many sponsorships or even any at all.
Compared to the work of Overney et al. \cite{Overney-rich}, this research brings us one step closer to understanding the incentive effect of sponsorship on individual developers by collecting feedback from participants in open source donations, taking the GitHub \textsc{Sponsor} as an example.

Although this article considers only the \textsc{Sponsor} mechanism, it lacks overall consideration and comparative analysis of all open source sponsorship platforms. However, we think that the article still provides some guidance in helping improve the mechanism itself and exploring the essence of open source donation.

This paper explored four aspects of the \textsc{Sponsor} mechanism: its who, what, why, and how. The main findings and insights are as follows.

\paragraph{Why do individuals participate or not in the \textsc{Sponsor} mechanism?}
Not all open source contributors endorse open source donation. There were more nonparticipants than participants.
Like the motivations for participation in traditional citizen science~\cite{domroese2017watch,larson2020diverse} and information-sharing crowdsourcing systems like Wikipedia~\cite{xu2015empirical}, developers are primarily intrinsically motivated to participate in open source contributions~\cite{open-source-get-paid}. However, because open source development activities are more complex and require significant maintenance, many contributors are looking for financial support~\cite{The-ethics,Schlueter-money,Steven-hard-work}.
Among the groups that support and use it, it is generally relationships built through the use of specific software that serve as the backbone of the sponsorship behavior.
In fact, many users want to reflect the difference between sponsors and nonsponsors in development activities and, in this way, change the method of open source collaboration and participation in open source donation. Such a change might not be very pleasant
and could lead to the open source sphere becoming money driven. We think that making the format personalized and configurable may meet the needs of more people without changing the nature of the open source sphere.

It is necessary for system designers to consider regional support and then make the \textsc{Sponsor} mechanism accessible and better for more people who want to participate by improving the user experience (\emph{e.g.,} better access to bill for tax).

\paragraph{How effective is sponsorship in motivating developer OSS activity?}
In a study of donations to projects, Overney et al.~\cite{Overney-rich} found that donation did not improve engineering activity. And in our study, we also found that
sponsorship has only a short-term positive stimulating effect on \emph{maintainers'} development activity. However, the impact does not last, and there is even a slight negative effect in the long term. A possible reason for this result is that most \emph{maintainers} do not receive sufficient sponsorship through the \textsc{Sponsor} mechanism to be motivated to contribute continuously.
This may reflect the characteristics of open source donations. The \emph{maintainer} passively receives sponsorship from the sponsor, and there is no compulsion for the act of sponsorship to occur. Thus, situations may arise that are similar to that of one of our questionnaire participants, who created heavily used tools but received no sponsorships. When compared horizontally with the results of other maintainers, such an outcome may have the negative effect of dealing a blow to maintainers and reducing their enthusiasm for making open source contributions.

For system designers, it is important to consider how to design conjunctive mechanisms, such as adding a ranking list according to the number of received or given sponsorships in the annual report or other locations. Therefore, the sponsorship mechanism can become a more continuous driving force, enhancing the impact of the sponsorship on developer activities.

\paragraph{Who is more likely to receive sponsorships?}
Participants' subjective perceptions conflict with the actual phenomenon. Participants believe that creating useful open source projects should lead to more sponsorships. However, we find that the most significant factor influencing the amount of sponsorship is social status. This inconsistent finding illustrates that participants want to express their gratitude or receive appreciation from others through the software usage relationship. However, it is not the case that those who develop sufficiently useful tools receive substantive sponsorship.
Given the feedback from participants in our questionnaire, this situation is likely to cause maintainers to complain about a lack of publicity for themselves or about the fact that their work leads to no more sponsorships.
At the same time, developers who make minor contributions to popular projects or outstanding contributions to niche projects may be ignored under this mechanism.
Comparing to project-oriented donation, \emph{e.g.,} open collective, patreon~\cite{Overney-rich}. Although the \textsc{Sponsor} mechanism is targeted at developers, which allows external contributors who do not own but are actively involved in popular projects to get donations. However, it is found through the results that \emph{sponsors} prefer project-oriented donation, \emph{i.e.,} the core developers or owners of popular or used projects are more likely to receive sponsorship.
Since some of the money donated to projects is spent on travel/food~\cite{Overney-rich}, we think it is needed to consider the percentage of contributors' contributions to achieve greater equity.

As for now, we think that for open source developers who want to get more sponsorship, it is essential to increase one's community visibility through advertising and help oneself get more attention by building open source projects that more people use.

\paragraph{What are the shortcomings of the \textsc{Sponsor} mechanism?}
The defects of the \textsc{Sponsor} mechanism are manifested in three main aspects: usage defects, object-oriented and support mechanisms, and personalization setting problems. At the same time, many developers believe that sponsorship behavior is not suitable for the open source ecosystem. The free nature of OSS leads to an unwillingness to pay. This finding shows that in addition to the problems with the mechanism itself, donations are not perfectly adapted to the open source ecosystem. The passivity, uncertainty, and instability inherent to donations make it difficult for \emph{maintainers} to rely on them and continue to make open source contributions for a long time. At the same time, the lack of reasonable evaluations of contributions and funding allocation makes it difficult for \emph{sponsors} to determine whom to sponsor and by how much.
So the bounty approach of "getting paid to do more" is recognized by some people than the donation approach, through which they can get paid immediately for the work and have more precise goals~\cite{Zhou-bountysource}. But how to balance the advantages of bounty and avoid regarding money as the guide of open source development may be the goal of future monetary incentive system design.
For more specific system design recommendations, see Section~\ref{Result-RQ4}.

Overall, the \textsc{Sponsor} mechanism is a good attempt and an essential step toward achieving reasonable and effective open source financial support.
As of now, the mechanism still needs further improvement to meet the needs of more developers.

\section{Threats to validity}
\label{threats}

For the questionnaire,
we did not do the detection of carelessly invalid responses~\cite{curran2016methods}.
First of all, the number of questions is small, the time required to answer is short, and there is no overlap between questions, so it is not feasible to judge the validity of the responses simply by the results. Secondly, we did not set attention check items to shorten user participation time. However, since users need to click on our questionnaire and jump to the SurveyMonkey site to respond after receiving the email, we think this has ensured the validity of the responses we received to some extent.
When conducting the second round of the questionnaire survey, to avoid disturbing participants excessively, we sent it only once. We did not send second or third reminder emails. At the same time, people who have not set up a \textsc{Sponsor} account may not care about the mechanism. As a result, the response rate was low.

For ITS analysis, data should be collected for the different factors for each time window. However, due to the lack of availability of timestamps in the GitHub API, some factors were measured only at their values at the time of data collection (\emph{e.g.}, \emph{in company}), as they do not change frequently.

For hurdle regression, the factors included in the models were several aspects related to the sponsorship of developers.
However, other factors may influence whether a developer can obtain sponsorship or how much funding is received.
Moreover, the number of sponsorships does not accurately indicate the amount of money that a developer receives from donations, as there exist different tiers and sponsors can withdraw their monthly sponsorship at any time. However, we do not have access to data on the actual donations received by each developer.
Developers may obtain donations from other platforms to maintain related projects. We did not consider all this funding in total or the activities of developers on other platforms.

This paper explored only the effectiveness of the \textsc{Sponsor} mechanism for individual users, but the \textsc{Sponsor} mechanism itself can also be used for organizational accounts. To avoid our analysis being confounded by the impact of such users, we processed our data accordingly. Therefore, the results do not apply to GitHub's organizational accounts.
According to statistics, 92\% of users who set up sponsors are individual users. 

\section{Conclusion and Future Work}
\label{conclusion}

This paper took GitHub's \textsc{Sponsor} mechanism as a case study and used a mixed qualitative and quantitative analysis method to investigate four dimensions of the mechanism.
Regarding why developers participate in the \textsc{Sponsor} mechanism, we found that it is mainly related to the use of OSS.
Regarding the mechanism's effectiveness, we found that the \textsc{Sponsor} system has only a short-term effect on development activities but that in the long term, there is a slight decrease.
We studied who obtains more sponsorships and found that the social status of the \emph{maintainer} in the community correlates most strongly with this outcome (the more followers, the more sponsorships a developer acquires).
Regarding the drawbacks of the mechanism, we found that in addition to the shortcomings in its use, participants felt that the \textsc{Sponsor} mechanism should better attract and support corporate sponsors. Some people thought that the open source donation method needed to be improved to attract more developers to participate.
Overall, we have explored the correlation between donation behavior and developers in open source communities using the GitHub \textsc{Sponsor} mechanism. In future work, we will further explore the following aspects: 1) the advantages and disadvantages of different open source donation platforms and the effectiveness of incentives for open source activities and 2) different types of open source financial support and the reasonableness and effectiveness of each mode.

\begin{acks}
This work is supported by China National Grand R\&D Plan (Grant No.2020AAA0103504).
Thanks to all GitHub users who response to the questionnaire.
\end{acks}

\bibliographystyle{ACM-Reference-Format}
\bibliography{sample-base}

\appendix
\section{Other platforms besides the \textsc{Sponsor} mechanism}
\begin{table}[htbp]
    \centering
    \footnotesize
    \renewcommand{\arraystretch}{1.3}
    \caption{Other platforms for obtaining OSS financial support}
    \label{donation-platforms}
    \begin{tabular}{p{0.34\linewidth} p{0.59\linewidth}}
    \toprule
    \textbf{Name} & \textbf{URL} \\
    \hline
    Bountysource & https://www.bountysource.com \\
    Flattr & https://flattr.com \\
    IssueHunt & https://issuehunt.io \\
    Kickstarter & https://www.kickstarter.com \\
    Liberapay & https://liberapay.com \\
    Gittip & https://gratipay.com \\
    Gratipay & https://gratipay.com \\
    OpenCollective & https://opencollective.com \\
    Otechie & https://otechie.com \\
    Patreon & https://www.patreon.com \\
    PayPal & https://www.paypal.com \\
    Tidelift & https://tidelift.com \\
    Tip4Commit & https://tip4commit.com \\
    LFX Mentorship (formerly CommunityBridge) & https://lfx.linuxfoundation.org/tools/mentorship \\
    Ko-fi & https://ko-fi.com \\
    \bottomrule
    \end{tabular}
\end{table}

\end{document}